\documentclass[a4paper,12pt]{article}
\usepackage{amsmath}
\usepackage{graphicx}

\begin{document}

\begin{center}

{\Large \bf Some implications of lepton flavor violating processes
in a supersymmetric Type II seesaw model at TeV scale}\\[20mm]

Raghavendra Srikanth Hundi\footnote{
The author's current address is: Centre for High Energy Physics,
Indian Institute of Science, Bangalore 560 012, India.
E-mail address: srikanth@cts.iisc.ernet.in}\\
Department of Theoretical Physics,\\
Indian Association for the Cultivation of Science,\\
2A $\&$ 2B Raja S.C. Mullick Road,\\
Kolkata - 700 032, India.\\[20mm]

\end{center}

\begin{abstract}
We have conceived a supersymmetric Type II seesaw model at TeV scale,
which has some additional particles consisting of scalar and fermionic triplet
Higgs states,
whose masses being around few hundred GeV. In this particular model, we
have studied constraints on the masses of triplet states arising from
the lepton flavor violating (LFV) processes, such as $\mu\to 3e$ and
$\mu\to e\gamma$. We have analyzed the implications of these constraints
on other observable quantities such as the muon anomalous magnetic moment
and the decay patterns of scalar triplet Higgses.
Scalar triplet Higgs states can decay into leptons and into supersymmetric
fields. We have found that the constraints from LFV can effect these various
decay modes.
\end{abstract}


\newpage

\section{Introduction}

The Standard Model (SM) has been a successful model and the only missing
piece of it is the Higgs boson. In the recent experiment at the Large Hadron
Collider (LHC), the discovery of
a Higgs-like particle has been reported \cite{LHC-Higgs}. As of now the discovery at the LHC
does not imply that it is a Higgs boson of the SM and it could even belong to the
physics beyond the SM. On going studies at the LHC will confirm this in future. As for
the physics beyond the SM, several motivations have been given \cite{BSM}. The important motivations
among these are the gauge hierarchy problem, smallness of neutrino masses, existence
of dark matter, etc. Although there is a growing belief in the physics beyond the SM,
the theoretical models in this category also have to deal with the constraints from the
flavor violating processes.
For a review on flavor violating processes, see Ref. \cite{fcnc-rev}.
The SM has been consistent with all the flavor violating processes
due to the Glashow-Iliopoulos-Maiani cancellation mechanism, and this cancellation
mechanism may not work in models of physics beyond the SM.

In this work, we have been motivated by arguments for physics beyond the SM
\cite{BSM}, especially related to
neutrino masses \cite{neut-rev}. Among the various models
for non-zero neutrino masses, Type II
seesaw mechanism offers a viable model \cite{Tp-II}. In this model, the scalar triplet Higgs with
hypercharge $Y=1$ can give Majorana masses to neutrinos by acquiring a vacuum expectation
value (vev) to the neutral component of the triplet Higgs.
Due to the seesaw mechanism \cite{Tp-II}, the vev of neutral triplet Higgs
can be as low as $\sim$ 1 eV, provided the masses of these states are
${\cal O}(10^{14})$ GeV.
As a result of this, for ${\cal O}(1)$ Yukawa couplings the neutrino mass
scale $m_\nu\sim$ 0.1 eV can be explained.
Supersymmetry (SUSY) \cite{susy,Martin}
has been proposed to solve the gauge hierarchy problem and it is one of
the main contenders for new physics.
To explore the models among the physics beyond the SM, supersymmetrizing the Type II seesaw
mechanism would be worth to do \cite{susyTp-II,Rossi}. In the supersymmetrized version of Type II
seesaw model, both the scalar and fermionic states of triplet Higgses will have super
heavy masses of ${\cal O}(10^{14})$ GeV. A positive aspect of having super heavy masses to
triplet Higgs states is that the lepton flavor violating (LFV) processes in both the non-SUSY and
SUSY versions of Type II seesaw model would be suppressed and
they can be within the
experimental limits. A negative point in these models is that these heavy
triplet states cannot be produced at the LHC, and hence a direct detection
is unlikely for the Type II seesaw mechanism.
For indirect signals of super heavy triplet states, see Ref. \cite{heavyTpII}.
Hence, for phenomenological studies at the LHC, we consider
a specific version of SUSY Type-II seesaw model, where we conceive
TeV scale masses for the triplet Higgs states.

In the Type II seesaw model, leptogenesis mechanism can be employed
to explain the asymmetry between matter and anti-matter
\cite{susyTp-II,lepgen}. In the non-SUSY version
of Type II seesaw model, the recent indication of LHC experiment on
the existence of Higgs boson \cite{LHC-Higgs} can also be accommodated \cite{hgamgam}.
In these models, the triplet Higgs states
can induce LFV processes such as $\mu\to 3e$, $\tau\to 3\mu$, $\tau\to e2\mu$, etc at tree level, and at 1-loop
level decays like $\mu\to e\gamma$, $\tau\to e\gamma$ and $\tau\to \mu\gamma$
can also happen.
None of the above mentioned LFV decay processes have been observed in
experiments
and stringent experimental upper bounds have been put on the decay 
branching ratios of these processes \cite{pdg}.
In fact, in the SUSY version of Type II seesaw model, the above mentioned
LFV processes can get additional radiative contributions which are induced
by slepton fields. These additional contributions due to slepton
fields also exist in the minimal supersymmetric standard
model (MSSM). In the MSSM, the off-diagonal elements in soft masses
of sleptons can generate LFV processes which are induced at 1-loop level.
As a result of this, constraints on model parameters may be
reduced in MSSM as compared to that in Type II seesaw model. Especially,
we may expect stringent bounds in Type II seesaw model from processes
such as $\mu\to 3e$ which take place at tree level.

In the literature, some work has already been done on the LFV processes in the non-SUSY
version of Type II seesaw model at TeV scale \cite{TpIIwk,Aker-etal,Fuku-etal}. Even in the SUSY version of
Type II seesaw model at TeV scale,
some work has been done in this direction \cite{Sena-Yama}. However, in Ref.~\cite{Sena-Yama},
a detailed study of
constraints on model parameters arising from LFV processes has
not been done.
Moreover, in Ref.~\cite{Sena-Yama}, the model has been motivated from high scale physics,
and due to renormalization group effects,
off-diagonal elements in the slepton mass matrices can become
non-zero at low energy scale. As a result of this, processes like
$\mu\to e\gamma$ can have additional contribution due to sletpon fields.

In this work, we have confined to the SUSY version of
Type II seesaw model at the low energy scale and assume zero off-diagonal elements in charged slepton and
sneutrino mass matrices. More precisely, we assume off-diagonal elements to be zero
in the soft mass-squared terms and also in the soft $A$-terms of the slepton fields.
This assumption  makes our work to be different from that in Ref.~\cite{Sena-Yama}.
Moreover, in our considered model, the LFV processes can
happen only due to the non-diagonal Yukawa couplings of triplet Higgs field with the lepton doublets.
Although we have neglected the contribution from slepton fields,
the LFV processes in our work are clearly different from that of non-SUSY version of
Type II seesaw model \cite{TpIIwk,Aker-etal,Fuku-etal}, since the fermionic partners
of scalar triplet fields will give additional contribution to the LFV processes
in our model.

The LFV processes in our model dominantly
depend on neutrino Yukawa couplings and masses of triplet Higgs states.
The Yukawa couplings can be determined from neutrino masses and mixing angles
as well as from vev of scalar triplet Higgs. We will show later that
the vev of scalar triplet Higgs can be around 1 eV in order to be compatible
with neutrino oscillation data. Hence, by determining the Yukawa couplings,
the experimental limits on LFV processes can put constraints on the masses
of triplet Higgs states.
We have studied implications of these constraints
on other observable quantities such as the muon anomalous magnetic
moment \cite{mug-2} and the decay patterns of scalar triplet Higgses.
Since the triplet states have TeV scale masses, they can be pair
produced at the LHC and their decay products give us experimental
signals of this model.
We have found that the constraints from LFV processes
can effect the decay channels of these fields.

The organization of our paper is as follows. In the next section we give a brief description
of SUSY version of Type-II seesaw model at TeV scale. In Sec. 3, we describe
various possible LFV processes in this model
and the expressions of their branching ratios.
In Sec. 4, we have presented
constraints due to the LFV processes on the masses of scalar and fermionic components of
triplet Higgs states of this model. In the same section, we have also given results on the contribution
of triplet states to the anomalous magnetic moment of the muon, $(g-2)_\mu$.
In Sec. 5, we have described
various decay channels of the scalar triplet Higgs states and their
branching ratios. In Sec. 6, we have commented on phenomenological signals of
this model in collider experiments. We conclude in
Sec. 7. We have given total scalar potential of this model in Appendix A.
Our conventions on neutralino and chargino mass matrices are described
in Appendix B.

\section{The Model}

The gauge symmetry of the SUSY Type II seesaw model is SU(3)$_C\times$SU(2)$_L\times$U(1)$_Y$.
The superpotential of this model can be written as \cite{susyTp-II,Rossi}
\begin{eqnarray}
W&=& W_{\rm MSSM} + W_{\rm II},
\nonumber \\
W_{\rm MSSM} &=& Y_u^{jk}Q_jH_uD^c_k + Y_d^{jk}Q_jH_dD^c_k + Y_e^{jk}L_jH_dE^c_k + \mu H_uH_d,
\nonumber \\
W_{\rm II} &=& Y_\nu^{jk}L_ji\sigma_2T_1L_k + \lambda_1H_di\sigma_2T_1H_d + \lambda_2H_ui\sigma_2T_2H_u
+ M{\rm Tr}(T_1T_2).
\label{E:WII}
\end{eqnarray}
In the above equation, $W_{\rm MSSM}$ is the superpotential of MSSM.
Here, $j,k=1,2,3$ are the family indices and $\sigma_j$ are Pauli matrices. $Q$ and $L$ are the quark
and lepton SU(2)$_L$ doublet superfields, respectively. $U^c$, $D^c$
and $E^c$ are SU(2)$_L$ singlet superfields which represent up-type,
down-type quarks and charged lepton, respectively.
$H_u$ and $H_d$ are the SU(2)$_L$ doublet superfields with hypercharges
$Y=\frac{1}{2}$ and $-\frac{1}{2}$, respectively.
$\mu$ and $M$ are the only two mass parameters in the above equation.
The terms in $W_{\rm II}$ contain SU(2)$_L$ triplet superfields
$T_1$ and $T_2$, whose hypercharges are
$Y=1$ and $-1$, respectively. The forms of $T_1$
and $T_2$ are
\begin{equation}
T_1=\left(\begin{array}{cc}
\frac{1}{\sqrt{2}}T_1^+ & T_1^{++} \\
T_1^0 & -\frac{1}{\sqrt{2}}T_1^+
\end{array}\right), \quad
T_2=\left(\begin{array}{cc}
\frac{1}{\sqrt{2}}T_2^- & T_2^0 \\
T_2^{--} & -\frac{1}{\sqrt{2}}T_2^-
\end{array}\right).
\end{equation}
The neutral part of scalar triplet Higgs in $T_1$ can acquire vev
and it generates masses to neutrinos. The expression for
neutrino mixing mass matrix is given below.
\begin{equation}
M_\nu^{jk} = 2Y_\nu^{jk}v^\prime_1,
\end{equation}
where $\langle \phi_1^0\rangle=v^\prime_1$. The appearance of factor 2 in
the above equation is due to
the Majorana nature of neutrino fields.
The vev of the neutral scalar part of $T_2$
is $\langle \phi_2^0\rangle=v^\prime_2$. From the naturalness of
parameters we can
expect $v^\prime_1\sim v^\prime_2$.
Here, we use the convention that the scalar parts of triplet Higgs
$T_1$ {\em are denoted by $\phi_1$s and their supersymmetric counter parts are denoted by $\Delta_1$s}.
We follow the similar convention to denote the scalar and fermionic
parts of $T_2$.
To generate realistic neutrino masses, we can choose $Y_\nu\sim {\cal O}(1)$ and $v^\prime_1\sim$
1 eV. By choosing $v_{1,2}^\prime\sim$ 1 eV, the upper limit ($\sim$ 1 GeV)
on the vevs of scalar triplet Higgses, which arises from precision electroweak
tests, can be satisfied.
The Yukawa couplings can be uniquely determined in terms of neutrino masses
and mixing
angles, whose relations in a matrix format can be written as
\begin{equation}
Y_\nu = \frac{1}{2v^\prime_1}U^*_{\rm PMNS}M_{\rm diag}U^\dagger_{\rm PMNS}, \quad
M_{\rm diag} = {\rm diag}(m_1,m_2,m_3),
\label{E:Yuk}
\end{equation}
where $m_{1,2,3}$ are the three neutrino mass eigenvalues and $U_{\rm PMNS}$ is the
Pontecorvo-Maki-Nakagawa-Sakata unitary matrix.
In the actual numerical analysis, we will see that either in the normal or inverted hierarchical mass pattern 
of neutrinos, the Yukawa couplings are $Y_\nu\sim 10^{-2}$ for $v^\prime_1\sim$ 1 eV. Unless there is 
a mechanism to justify the order 2 suppression in $Y_\nu$, we may take this as a natural value in the 
Type II seesaw mechanism.

In the non-SUSY version of Type II seesaw model, non-zero vev of scalar
triplet Higgs arises due to the
tri-linear coupling between triplet and doublet Higgs states
\cite{TpIIwk,Aker-etal}. In the
SUSY version of Type II seesaw model, this tri-linear coupling is equivalent
to the $\lambda_{1,2}$-terms of Eq. (\ref{E:WII}).
To realize the possibility of $v^\prime_{1,2}\sim$ 1 eV, we can take the dimensionless
parameters $\lambda_{1,2}\sim {\cal O}(1)$ and the fermionic triplet Higgs
mass $M\sim 10^{14}$ GeV
\cite{susyTp-II,Rossi}. However, as explained before, in this case the masses of scalar and
fermionic triplet Higgs states will be super heavy and a direct search for them at colliders is
unlikely. Alternatively, we can consider another possibility where $\lambda_{1,2}\sim 10^{-10}$ and $M\sim$
1 TeV \cite{Sena-Yama}. In this later case, triplet Higgs states can be accessible
at the on-going LHC experiment.
The justification for the suppression of dimensionless couplings can be given if we embed the model in a high
scale theory like supergravity \cite{Sena-Yama}. Supergravity is a realistic
scenario where supersymmetry breaking can be achieved through some gauge singlet
fields known as hidden sector fields ($X$) \cite{susy,Martin}. Hidden sector fields can break
supersymmetry at an intermediate energy scale $\langle X\rangle = \Lambda\sim 10^{11}$ GeV,
and as a result, the
supersymmetric fields will have masses of the order of $\frac{\Lambda^2}{M_P}\sim$ 1 TeV.
Here, $M_P$ is the Planck
scale which is $\sim 10^{19}$ GeV. By embedding the model in a high scale theory,
we may identify $\lambda_{1,2}\sim Y_\nu\frac{\langle X\rangle}{M_P}$, which gives
the necessary suppression in the $\lambda_{1,2}$.
Although a realistic construction for the above model can be made
by embedding it in a high scale theory,\footnote{See Ref. \cite{HPT}, for embedding of another variety of SUSY
model in a supergravity setup.} it is beyond the scope of this work.
We here, on the phenomenological grounds,
consider a low energy setup of the above described SUSY Type II seesaw model.

The implications of supersymmetry breaking is to generate soft terms of scalar potential
in the low energy regime.
For full form of the scalar potential of this
model, see Appendix A. Below we have given soft terms in the
scalar potential which contain scalar triplet Higgs states.
\begin{eqnarray}
V^{\rm triplet}_{\rm soft} &=& m_{\phi_1}^2Tr(\Phi_1^\dagger\Phi_1) + m_{\phi_2}^2Tr(\Phi_2^\dagger\Phi_2) +
[B_TMTr(\Phi_1\Phi_2) + (A_\nu Y_\nu)^{jk}\tilde{L}_ji\sigma_2\Phi_1\tilde{L}_k
\nonumber \\
&&+ (A_{\lambda_1}\lambda_1)
H_di\sigma_2\Phi_1H_d + (A_{\lambda_2}\lambda_2)H_ui\sigma_2\Phi_2H_u + {\rm h.c.}],
\label{E:vsoft}
\end{eqnarray}
where the form of $\Phi_{1,2}$ is same as that of $T_{1,2}$ with its superfields being replaced
by their scalar components. All the various mass parameters in the above equation would be at around 1 TeV.
The term $Tr(\Phi_1\Phi_2)$ gives mixing masses between the components of
$\Phi_1$ and $\Phi_2$. In
the following basis: $\psi_{++}=(\phi_1^{++},\left(\phi_2^{--}\right)^*)^{\rm T}$,
$\psi_{+}=(\phi_1^{+},\left(\phi_2^{-}\right)^*)^{\rm T}$,
$\psi_{0}=(\phi_1^{0},\left(\phi_2^{0}\right)^*)^{\rm T}$, the mixing mass-squared
terms of doubly charged, singly charged and neutral scalars can be written as
\begin{equation}
\psi_{++}^\dagger M_{++}^2\psi_{++},\quad
\psi_{+}^\dagger M_{+}^2\psi_+, \quad
\psi_0^\dagger M_{0}^2\psi_0.
\end{equation}
The form of $M_{++}^2$ is
\begin{equation}
M_{++}^2 = \left(\begin{array}{cc}
M^2+m_{\phi_1}^2+m_{++}^2 & \left(B_TM\right)^* \\
B_TM & M^2+m_{\phi_2}^2-m_{++}^2 \end{array}\right),
\label{E:M2pp}
\end{equation}
where $m_{++}^2= \frac{g^2-g^{\prime 2}}{2}\cos(2\beta)v^2$, which arises
due to $D$-terms in the SUSY scalar potential (see Appendix A).
By replacing
$m_{++}^2$ with $m_+^2= -\frac{g^{\prime 2}}{2}\cos(2\beta)v^2$ and
$m_0^2= -\frac{g^2+g^{\prime 2}}{2}\cos(2\beta)v^2$ in $M_{++}^2$, we
get the corresponding forms for $M_+^2$
and $M_0^2$, respectively. Here, we have taken
the electroweak scale as $v=174$ GeV and $\beta$ is defined as $\tan\beta=
\langle H_u^0\rangle /\langle H_d^0\rangle$.
$g,g^\prime$ are the gauge couplings of SU(2)$_L$
and U(1)$_Y$ gauge groups, respectively.
Since the above mixing mass matrices ($M_{++}^2$, $M_{+}^2$, $M_0^2$) are hermitian,
they can be diagonalized by unitary matrices which we denote by $U^{++}$,
$U^+$ and $U^0$, respectively. For real parameters in the soft scalar
potential, we can express $U^{++}$ in terms of model parameters, which is
given below.
\begin{equation}
U^{++}=\left(\begin{array}{cc}
\cos\theta^{++} & -\sin\theta^{++} \\
\sin\theta^{++} & \cos\theta^{++}
\end{array}\right),\quad
\sin 2\theta^{++}=\frac{2B_TM}{\sqrt{(m_{\phi_1}^2-m_{\phi_2}^2+2m_{++}^2)^2+4(B_TM)^2}}.
\label{E:U++}
\end{equation}
Here, $\theta^{++}$ is the mixing angle between $\phi_1^{++}$ and
$\left(\phi_2^{--}\right)^*$.
Analogously, the elements of matrices $U^+$, $U^0$ can also be expressed in terms
of model parameters.

Before concluding this section, we comment on the masses of fermionic
triplet Higgs states. From the last term of $W_{\rm II}$,
Eq. (\ref{E:WII}), we can see that the dominant
contribution to the masses of these fields is $M$. However, after electroweak symmetry
breaking, fermionic fields like $\Delta^+_{1}$, $\Delta^-_{2}$ and $\Delta^0_{1,2}$ will have
some mixing masses with higgsinos, winos and bino. Because of this mixing,
the neutralino and chargino mass matrices \cite{susy,Martin} of MSSM will
be extended to 6$\times$6 and 3$\times$3, respectively, in this model.
The mixing masses can happen
due to $\lambda_{1,2}$-terms of $W_{\rm II}$ and also due to gauge invariant
kinetic $D$-terms of $T_{1,2}$ (See Sec. 5 for $D$-terms of $T_1$ and $D$-terms of $T_2$
can be analogously written).
The corrections due to former terms are negligible due to
the suppressed values of $\lambda_{1,2}$. The $D$-terms also give negligible corrections
because these mixing masses are proportional to $v_{1,2}^\prime$.
Since these corrections are $\sim$ 1 eV, we can safely take all the fermionic
triplet Higgs states to be degenerate with a mass of $M$. As a result
of this, in
this work, we have taken both the neutralino and chargino mass matrices to
be 4$\times$4 and 2$\times$2, respectively, which are described in Appendix B.

\section{LFV processes}

As described in Sec. 1, in our model, we assume vanishingly small off-diagonal
elements in the soft mass-squared terms and also in $A$-terms of
the slepton fields. As a result of this, in
the lepton sector of our model,
the Yukawa couplings in the first term of $W_{\rm II}$,
Eq. (\ref{E:WII}), can only generate flavor
changing processes, whose interaction terms in the Lagrangian are given below.
\begin{equation}
{\cal L} = Y_\nu^{jk}\left[-2\nu_L^j\Delta_1^0\tilde{\nu}_L^k - \nu_L^j\nu_L^k\phi_1^0
+\sqrt{2}\left(\nu_L^j\Delta^+_1\tilde{e}^k_L + e^j_L\Delta^+_1\tilde{\nu}_L^k + \nu_L^je_L^k\phi^+_1\right)
+2e_L^j\Delta^{++}_1\tilde{e}_L^k + e_L^je_L^k\phi^{++}_1\right] + {\rm h.c.}
\label{E:lagII}
\end{equation}
The last four terms in the above equation can drive LFV processes
at tree level and
and also at 1-loop level. Below we have described these processes.

\subsection{LFV processes at tree level}

The off-diagonal elements in the last term of Eq. (\ref{E:lagII})
generate LFV processes at tree
level such as $\mu^-\to e^+e^-e^-$, $\tau^-\to e^+e^-e^-$, $\tau^-\to\mu^+e^-\mu^-$,
$\tau^-\to e^+\mu^-\mu^-$, $\tau^-\to e^+e^-\mu^-$, $\tau^-\to\mu^+e^-e^-$, $\tau^-\to\mu^+\mu^-\mu^-$.
The experimental upper limit on $BR(\mu^-\to e^+e^-e^-)$
is $10^{-12}$ \cite{pdg} and the corresponding upper limits on the
branching ratios of $\tau$-decays are about $\sim 10^{-8}$ \cite{pdg}.
These LFV processes are driven
by the scalar field $\phi_1^{++}$. As explained in the previous section,
in this model there
is a mixing between $\phi_1^{++}$ and $\phi_2^{--}$. Hence,
the contributions due to both these fields
should be summed in the amplitudes of these processes.
Below we have given expressions for
branching ratios of the above mentioned decays.
\begin{eqnarray}
BR(\mu^-\to e^+e^-e^-) &=& \frac{8|Y_\nu^{12}|^2|Y_\nu^{11}|^2}{g^4}m_W^4
\left[\frac{|U^{++}_{11}|^2}{m_{\phi^{++}_1}^2}+\frac{|U^{++}_{12}|^2}{m_{\phi^{++}_2}^2}\right]^2,
\nonumber \\
BR(\tau^-\to \ell_j^+\ell_m^-\ell_l^-) &=& S\frac{16|Y_\nu^{j3}|^2|Y_\nu^{lm}|^2}{g^4}m_W^4
\left[\frac{|U^{++}_{11}|^2}{m_{\phi^{++}_1}^2}+\frac{|U^{++}_{12}|^2}{m_{\phi^{++}_2}^2}\right]^2
BR(\tau^-\to\mu\bar{\nu}_\mu\nu_\tau),
\nonumber \\
\end{eqnarray}
where $m_W$, $m_{\phi^{++}_{1,2}}$ are the masses of $W$-boson and doubly charged $\phi$-fields,
respectively. Here, $\ell_1=e$ and $\ell_2=\mu$ (Here, the muon field ($\mu$)
is different from $\mu$-parameter of Eq. (\ref{E:WII})).
$S$ is a symmetric factor which equals to
$\frac{1}{2}$ if $l=m$, otherwise it equals to 1. The branching
ratio of $\tau^-\to\mu\bar{\nu}_\mu\nu_\tau$ is $\approx 0.17$. The elements
of $Y_\nu$ can be computed from Eq. (\ref{E:Yuk}) by knowing
the neutrino masses and mixing angles. The values of $U_{11}^{++}$ and
$U_{12}^{++}$ can be computed from model parameters through Eq. (\ref{E:U++}),
for real soft mass parameters.

In the previous section, we have motivated our model in such a way
that in order to explain the smallness of neutrino
masses, a natural parameter space is $v_1^\prime\sim$ 1 eV so that the
elements of $Y_\nu$ are nearly unsuppressed. Hence, for this choice
of parameter space, the above mentioned LFV processes can give lower
bounds on the masses of doubly charged scalar fields.
As explained before that due to similarity in the form of matrices $M_{++}^2, M_+^2, M_0^2$, the
above mentioned bounds on the doubly charged fields will translate into similar lower bounds on
the masses of singly charged and neutral scalar triplet fields.
Hence, we can conclude that in our scenario the LFV processes at tree level
can constrain the masses of scalar components of the triplet states.

\subsection{Radiative LFV processes}

The last four terms of Eq. (\ref{E:lagII}) can generate LFV processes
at 1-loop level. These are $\mu\to e\gamma$, $\tau\to e\gamma$ and
$\tau\to\mu\gamma$. The experimental upper limit on $BR(\mu\to e\gamma)$ is
$2.4\times 10^{-12}$ at 90$\%$ C.L. \cite{MEG},
and the corresponding upper limits on $BR(\tau\to e\gamma,\mu\gamma)$
are about $10^{-8}$ \cite{pdg}.
We will show later that the upper bounds on the branching ratios of
radiative LFV processes can put lower bounds on the
masses of fermionic triplet Higgs states.

Let us consider the decay process $\ell_j(p)\to\ell_i(p^\prime)+\gamma(q)$, which takes
place at one loop level. Here, $i,j=1,2,3$ are family indices. $\ell_i$ and $\ell_j$ are some
negatively charged leptons with 4-momenta $p^\prime$ and
$p$, respectively. The outgoing $\gamma$ has 4-momenta $q=p-p^\prime$.
Below we present the decay width for $\ell_j(p)\to\ell_i(p^\prime)+\gamma(q)$,
where we have
neglected the left-right mixing of charged sleptons.
The decay width of the above
process is governed by the amplitude which has the following form,
where there is no summation on the indices $i,j$.
\begin{equation}
i{\cal M}=ie\bar{u}_i(p^\prime)\left[A^{ij}_R\frac{1+\gamma_5}{2}+A^{ij}_L\frac{1-\gamma_5}{2}\right]i\sigma^{\mu\nu}q_\nu
\epsilon_\mu^*(q) u_j(p).
\end{equation}
Here, $u_i$ and $u_j$ are the Dirac spinors of the charge leptons $\ell_i$ and $\ell_j$, respectively, and
$\epsilon_\mu(q)$ is the polarization of photon.
The forms of $A^{ij}_R$ and $A^{ij}_L$ are given below, where there is no summation
on the indices $i,j$.
\begin{eqnarray}
&& A^{ij}_R = A_{ij}m_{\ell_j},\quad A^{ij}_L = A_{ij}m_{\ell_i},
\nonumber \\
&& A_{ij}=\sum_{k=1}^3\left\{-\frac{\left(Y_\nu^{ki}\right)^*Y_\nu^{kj}}{12\pi^2}
\left[\left(\frac{|U^{++}_{11}|^2}{m_{\phi^{++}_1}^2}+\frac{|U^{++}_{12}|^2}{m_{\phi^{++}_2}^2}\right)
+\frac{1}{8}\left(\frac{|U^{+}_{11}|^2}{m_{\phi^{+}_1}^2}+\frac{|U^{+}_{12}|^2}{m_{\phi^{+}_2}^2}\right)\right] 
\right. \nonumber \\
&& \left. + \frac{\left(Y_\nu^{ki}\right)^*Y_\nu^{kj}}{16\pi^2M^2}
\left[2f_1(x_k^{++})+4f_2(x_k^{++})+f_2(x_k^{+})\right]\right\},
\label{E:ampli} \\
&& x_k^{++}=\frac{m_{\tilde{l}_k}^2}{M^2},\quad x_k^{+}=\frac{m_{\tilde{\nu}_k}^2}{M^2},
\nonumber \\
&& f_1(x)=\frac{1}{(1-x)^4}\left[\frac{1}{3}+\frac{x}{2}-x^2+\frac{x^3}{6}+x\log(x)\right],
\nonumber \\
&& f_2(x)=\frac{1}{(1-x)^4}\left[\frac{1}{6}-x+\frac{x^2}{2}+\frac{x^3}{3}-x^2\log(x)\right].
\end{eqnarray}
Here, $m_{\ell_i}$, $m_{\tilde{l}_k}$ and $m_{\tilde{\nu}_k}$ are the masses of
charged lepton, charged slepton and sneutrino fields, respectively.
The decay width of $\mu\to e\gamma$ is given by
\begin{equation}
\Gamma(\mu\to e\gamma)=\frac{e^2}{16\pi}\left(|A^{12}_R|^2+|A^{12}_L|^2\right)\frac{(m_\mu^2 - m_e^2)^3}{m_\mu^3}.
\end{equation}
After neglecting the electron mass, the branching ratio of $\mu\to e\gamma$
is
\begin{equation}
Br(\mu\to e\gamma) = \frac{\Gamma(\mu\to e\gamma)}{\Gamma(\mu\to e\bar{\nu}_e\nu_\mu)}
= \frac{48\alpha\pi^3}{G_F^2}\left(A_{12}\right)^2,
\end{equation}
where $\alpha=\frac{e^2}{4\pi}$ and $G_F=1.166\times 10^{-5}$ GeV$^{-2}$.
The branching ratio of $\tau\to e\gamma$ can be computed from
\begin{equation}
Br(\tau\to e\gamma) = \frac{48\alpha\pi^3}{G_F^2}\left(A_{13}\right)^2
BR(\tau\to\mu\bar{\nu}_\mu\nu_\tau).
\end{equation}
In the above expression by replacing $A_{13}\to A_{23}$, we can get the expression for
branching ratio of $\tau\to\mu\gamma$. In these expressions we have applied
the approximation $m_\mu^2\ll m_\tau^2$.

The expression for the muon anomalous magnetic moment, $(g-2)_\mu$ \cite{mug-2},
can be found from the same amplitude of $\ell_j\to\ell_i+\gamma$, which
is described above. By identifying $\ell_i=\ell_j=\mu$, the necessary amplitude for
the $(g-2)_\mu$ can be written as
\begin{equation}
i{\cal M} = ie\bar{u}_\mu(p^\prime)\left[A^{22}_R\frac{1+\gamma_5}{2}+A^{22}_L
\frac{1-\gamma_5}{2}\right]i\sigma^{\mu\nu}q_\nu
\epsilon^*_\mu(q) u_\mu(p),
\end{equation}
where $u_\mu$ is the Dirac spinor of the muon. From the above amplitude, we can read
the contribution to the $(g-2)_\mu$ due to the triplet Higgs states, whose
expression is given below.
\begin{equation}
\Delta a^{\rm T}_\mu = \left(A^{22}_R+A^{22}_L\right)2m_\mu = 2A_{22}m_\mu^2.
\label{E:g-2}
\end{equation}

Here we comment on our results on the decay branching ratios of flavor
changing processes with the previously work done in the non-SUSY \cite{Aker-etal}
and SUSY \cite{Sena-Yama} versions of the Type II seesaw model. The LFV
processes at tree level are driven by the doubly charged
scalar triplet fields.
In the limit $B_T=0$, the mixing between the fields $\phi^{++}_1,\phi^{--}_2$
will vanish and the
branching ratios of these processes reduce to the expressions
as they are given in Ref. \cite{Aker-etal}. The amplitudes
for radiative decay processes, such as $\mu\to e\gamma$,
get contribution from scalar (1st line of Eq. (\ref{E:ampli})) as well as
from fermionic (2nd line of Eq. (\ref{E:ampli})) components
of triplet Higgs. Again, in the limit $B_T=0$, the contribution from
first line of Eq. (\ref{E:ampli}) reduces to the expression as it is given
in Ref. \cite{Aker-etal},
while the fermionic triplet contribution of Eq. (\ref{E:ampli}) has a similar
form to the corresponding expression given in Ref. \cite{Sena-Yama}.
However, the sign
proportional to the $f_1(x^{++}_k)$-term
is given with a minus sign in Ref. \cite{Sena-Yama}.

\section{Constraints from the LFV processes}

Before explaining constraints from the LFV processes, we here make brief
comments on relaxing constraints from
the tree level LFV processes. Among these,
we can expect stringent limits from $BR(\mu\to 3e)$. To
suppress limits from $BR(\mu\to 3e)$, we can fine tune the Yukawa couplings
$Y_\nu^{12},Y_\nu^{11}$ to be vanishingly small \cite{TpIIwk,Aker-etal,Fuku-etal}.
However, it has been reported in Ref. \cite{Chak-etal} that to achieve
$Y_\nu^{12}=0$, the neutrino mixing angle $\theta_{13}$ will have to be
too small which is not
consistent with the recently measured value of $\theta_{13}$ at the
Double Chooz, Daya Bay and RENO experiments \cite{theta13}.
Nevertheless, here our motivation is that we
choose generic values for neutrino masses and mixing angles, and study
bounds on the masses of triplet Higgs states.

The six neutrino Yukawa couplings in this model, Eq. (\ref{E:Yuk}),
are determined
by the neutrino masses and mixing angles. The mixing angles are incorporated
in the unitary matrix $U_{\rm PMNS}$, and we have parametrized this matrix
according to the convention in Ref. \cite{pdg}. Here, without loss of
generality, we
have chosen the CP violating phase $\delta$ and the two Majorana phases
to be zero. We have taken
the neutrino mass-squared differences as \cite{glob-fit}:
$m_{\rm solar}^2=m_2^2-m_1^2=7.62\times10^{-5}~{\rm eV}^2$ and
$m_{\rm atm}^2=m_3^2-m_1^2=2.53(-2.4)\times10^{-3}~{\rm eV}^2$. Here, the
term in bracket gives inverted hierarchical mass pattern for neutrinos.
To be consistent with the above neutrino mass-squared values,
we can choose three different hierarchical mass patterns,
which are described below.
\begin{eqnarray}
&&{\rm Normal~ hierarchy~(NH)}:m_1=0,\quad m_2=m_{\rm solar},\quad m_3=m_{\rm atm}
\nonumber \\
&&{\rm Inverted~ hierarchy~(IH)}:m_3=0,\quad m_1=m_{\rm atm},\quad m_2=\sqrt{m_{\rm solar}^2+m_1^2}
\nonumber \\
&&{\rm Degenerate~ Neutrinos~(DN)}:m_1=0.3~{\rm eV},\quad m_2=\sqrt{m_{\rm solar}^2+m_1^2},
\quad m_3=\sqrt{m_{\rm atm}^2+m_1^2}
\nonumber \\
\end{eqnarray}
As for the mixing angles, we have taken them as: $\sin\theta_{12}=\frac{1}{\sqrt{3}}$,
$\sin\theta_{23}=\frac{1}{\sqrt{2}}$ and $\sin\theta_{13}=0.1737$. Here
$\theta_{13}=10^{\rm o}$ and the other two angles are fitted to the
tri-bimaximal values \cite{tribi}. All these values are consistent with the
global fitting to the neutrino oscillation data, done in Ref. \cite{glob-fit}.

After determining the Yukawa couplings, $BR(\mu\to 3e)$ can put limits
on $m_{\phi^{++}_1}$ and $m_{\phi^{++}_2}$. However in this
analysis, we also have to know the values of $U^{++}_{11},U^{++}_{12}$.
It can be seen from Eq. (\ref{E:U++}) that
for generic SUSY parameter space, where $B_TM\sim M^2\sim m^2_{\phi_{1,2}}$,
$U^{++}_{11},U^{++}_{12}\sim{\cal O}(1)$. Hence the lower bound on
$m_{\phi^{++}_1}$ would be nearly the same as on $m_{\phi^{++}_2}$.
Alternatively,
to simplify this task, we may choose the soft parameters $B_T=0$ and
$m_{\phi_1}^2\sim m_{\phi_2}^2$. In this
case, $\phi^{++}_1$ and $\phi^{--}_2$ will be decoupled away from each other and we get
lower bound on $m_{\phi^{++}_1}$ from $BR(\mu\to 3e)$.
From Eq. (\ref{E:M2pp}), it can be noticed that for $\tan\beta\sim 10$,
the electroweak corrections to the triplet Higgses would be at most
$\sim$10 GeV. Hence, the lower bound
on $m_{\phi^{++}_1}$ will put nearly the same lower bound on
$m_{\phi^{++}_2}$. In fact,
the arguments given below Eq. (\ref{E:M2pp}) would suggest that similar
amount of lower bounds will apply on the singly charged and neutral
triplet scalar fields. Hence from the above argument of simplicity
we choose $B_T=0$ in this section.

In Tab. 1 we have presented lower bounds on the mass of $\phi^{++}_1$ which
arise from $BR(\mu\to 3e)<10^{-12}$.
\begin{table}[!h]
\begin{center}
\begin{tabular}{||c|c|c|c||} \hline
 & NH & IH & DN \\ \hline
$v^\prime_1$ & $m_{\phi^{++}_1}$ & $m_{\phi^{++}_1}$ & $m_{\phi^{++}_1}$ \\
1.0 eV & 631.8 GeV & 1.71 TeV & 1.32 TeV \\
0.5 eV & 1.26 TeV & 3.41 TeV & 2.64 TeV \\
0.1 eV & 6.32 TeV & 17.07 TeV & 13.21 TeV \\ \hline
\end{tabular}
\end{center}
\caption{Lower bounds on the mass of $\phi^{++}_1$ arising
from $BR(\mu\to 3e)<10^{-12}$,
for different values of $v^\prime_1$. These lower bounds are given in
all the three hierarchical mass patterns of neutrinos.}
\end{table}
We have checked that the lower bounds on $m_{\phi^{++}_1}$ due to
$BR(\mu\to 3e)<10^{-12}$ will simultaneously satisfy the
experimental limits on the branching ratios of $\tau$ decays
such as $\tau\to 3e$, $\tau\to e2\mu$, etc. The lower bounds in Tab. 1
can be compared to the lower bound of about 400 GeV on $m_{\phi^{++}}$ by
the CMS collaboration of the LHC experiment \cite{CMS}. From Tab. 1, we can notice
that the lower bounds in the case of NH are much lower
compared to that in IH and DN cases. The product $Y_\nu^{12}\times Y_\nu^{11}$,
which determines $BR(\mu\to 3e)$, is lower in the case of NH as compared to
that in IH and DN cases. However, if we look at numerical values,
for $v^\prime_1=$ 1.0 eV,
in both the NH and IH cases the elements of $Y_\nu$ are
$\sim 10^{-3}$, whereas, in the case of DN
the diagonal and off-diagonal elements of $Y_\nu$ are around 0.1
and $10^{-4}$ respectively. The lower bound on $m_{\phi^{++}_1}$ increases
with decreasing $v^\prime_1$, since from Eq. (\ref{E:Yuk}) we
see that $Y_\nu\sim\frac{1}{v^\prime_1}$. In fact, from Tab. 1, for
$v^\prime_1=$ 0.1 eV, the masses of scalar triplets are so high that there
is very less chance of their detection at the current LHC experiment.

Now, by inputting the lower bounds of the masses of scalar triplet
Higgses in the radiative LFV processes, such as
$\mu\to e\gamma$, we can derive lower bounds on the masses of fermionic
triplet Higgs states. From the expressions of decay branching
ratios of $\ell_j\to\ell_i\gamma$, which are given in the previous section,
we can notice that the masses of charged slepton and sneutrino fields will also
contribute to these radiative processes.
For simplicity, we have chosen degenerate masses for the
three charged sleptons ($m_{\tilde{l}}$) and for the three sneutrino fields
($m_{\tilde{\nu}}$). Regarding the masses of
scalar components of triplet
Higgses, as explained previously, the electroweak corrections can be
at most $\sim$10 GeV, and so in our numerical analysis we have taken
$m_{\phi^{++}_1}\approx m_{\phi^{+}_1}$. Moreover, in our
analysis, we have fixed the values of $m_{\phi^{++}_1}$
to the lower limits as they are given in Tab. 1, and we comment below on
what may
happen if we increase its value.
From the experimental limits on radiative LFV decays \cite{pdg,MEG},
we expect stringent constraints on model parameters from
$BR(\mu\to e\gamma)$. As a result of this, in the analysis,
for some fixed values of $m_{\tilde l}$ and $m_{\tilde\nu}$, we
first check if the constraints from $BR(\tau\to e\gamma,\mu\gamma)$
are satisfied and then compute
$BR(\mu\to e\gamma)$ as a function of fermionic triplet Higgs mass, $M$.

In Fig. 1, in the case of NH,
we have given constraints on $M$ from the above mentioned
radiative LFV processes.
\begin{figure}[!h]
\begin{center}

\includegraphics[height=2.5in, width=2.5in]{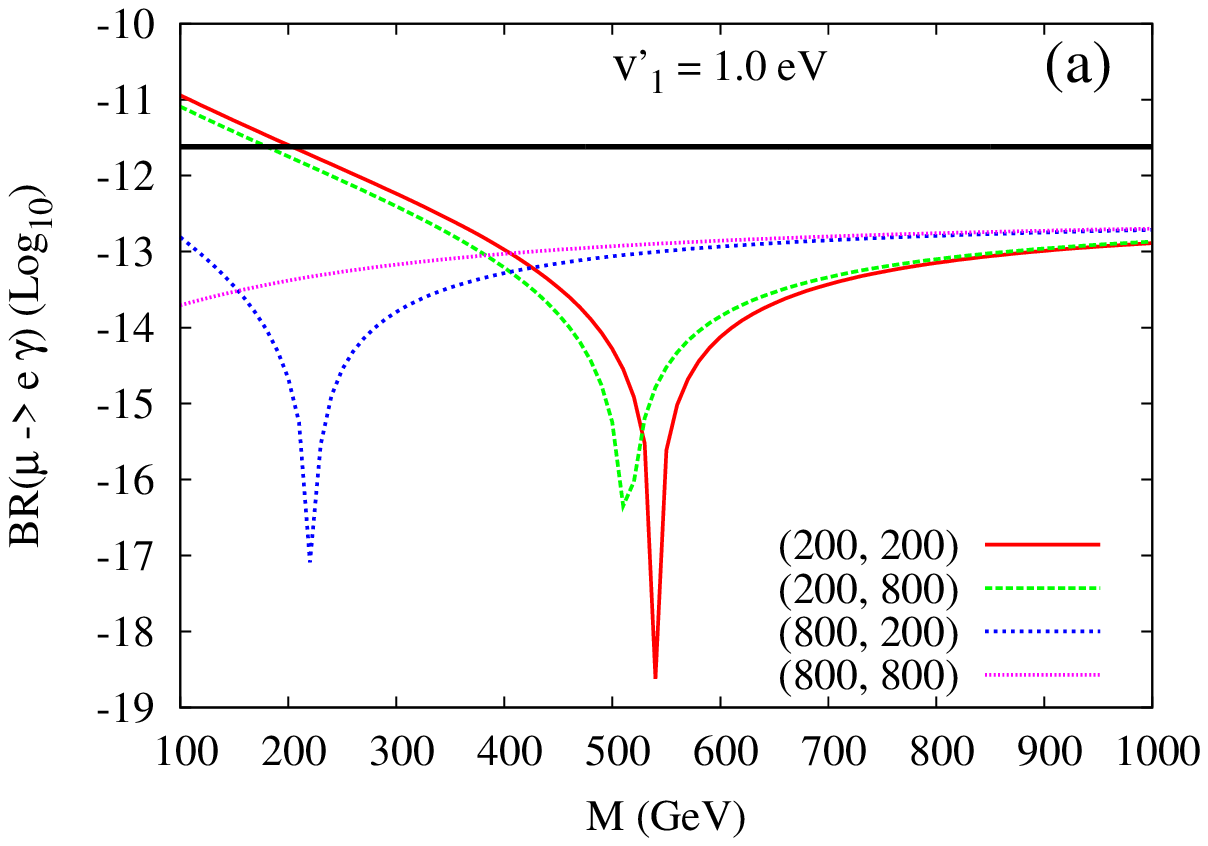}
\includegraphics[height=2.5in, width=2.5in]{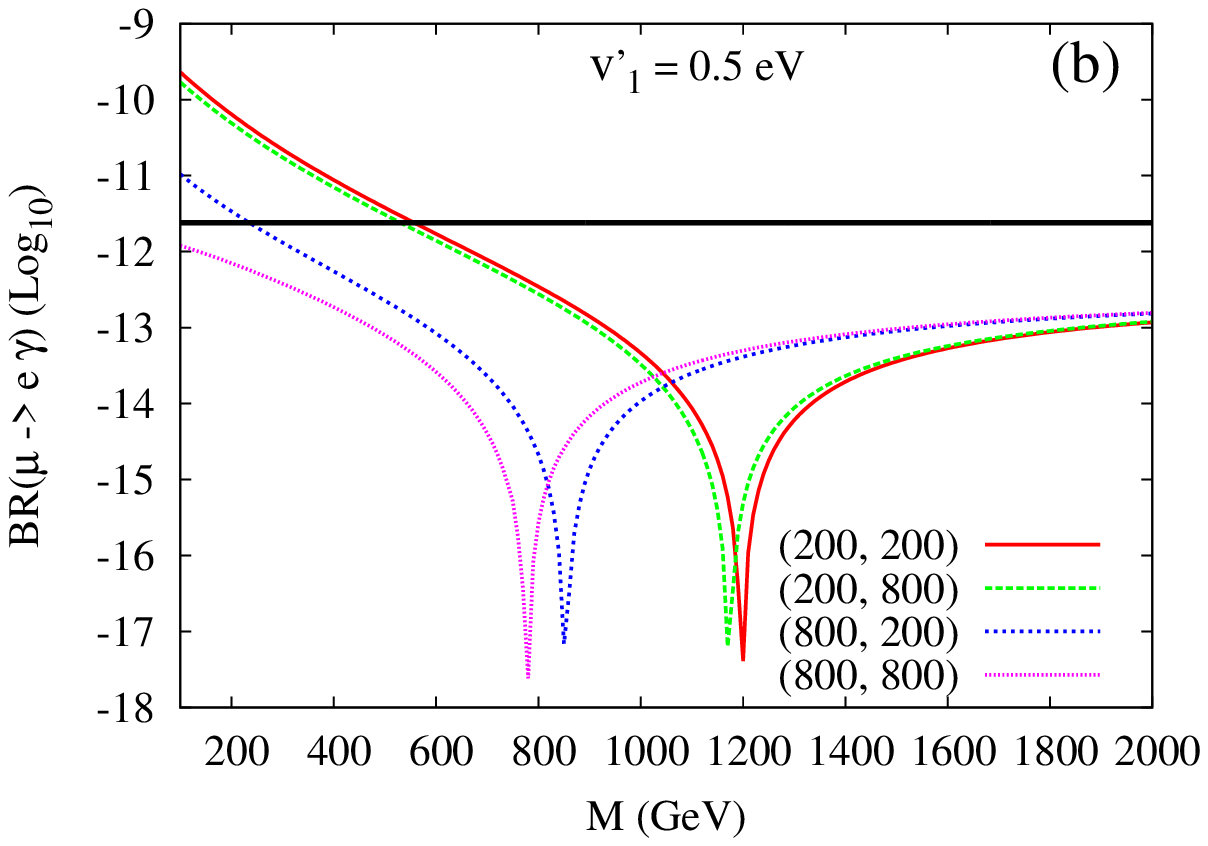}

\includegraphics[height=2.5in, width=2.5in]{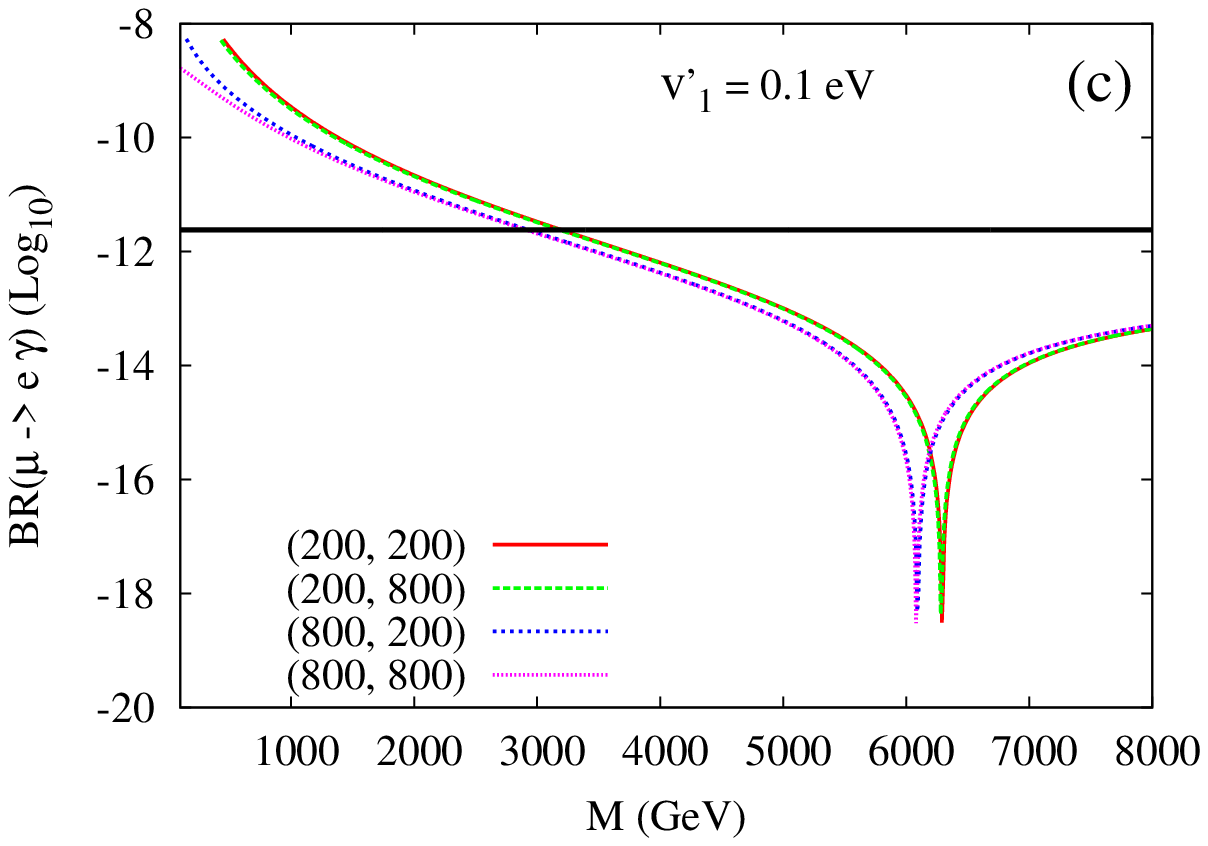}

\end{center}
\vspace*{-1cm}
\caption{In the normal hierarchy, $\log_{10}(BR(\mu\to e\gamma))$ has been plotted against
the mass of fermionic triplet Higgs. The three plots are for $v^\prime_1 =$ 1.0, 0.5 and 0.1 eV.
In each of these plots, the four lines are for different values
of charged slepton and sneutrino masses, which are represented in the format $(m_{\tilde l},m_{\tilde \nu})$
in GeV units.
The horizontal line in these plots indicate $BR(\mu\to e\gamma)=
2.4\times 10^{-12}$, and the area below this line is allowed.
The lower limit on the x-axis is 100 GeV.}
\end{figure}
In Fig. 1(a), we have fixed $v^\prime_1=1.0$ eV and plotted $BR(\mu\to e\gamma)$
versus $M$ for four different combinations of $(m_{\tilde l},m_{\tilde \nu})$.
Among these four different combinations, $(m_{\tilde l},m_{\tilde \nu})$ =
(200 GeV, 200 GeV) has given stringent lower limit on $M$ which is about
200 GeV. The next stringent limit on $M$ has come from the other combination
of $(m_{\tilde l},m_{\tilde \nu})$ = (200 GeV, 800 GeV), which sets $M\geq$ 180 GeV.
Whereas, the other two combinations such as $(m_{\tilde l},m_{\tilde \nu})$ =
(800 GeV, 200 GeV) and (800 GeV, 800 GeV) have put no limits on $M$. In
Figs. 1(b) and 1(c), we have decreased $v^\prime_1$ to 0.5 eV and 0.1 eV,
respectively. In these two plots we can observe that the lower limits on
$M$ will be stringent from the combination $(m_{\tilde l},m_{\tilde \nu})$ =
(200 GeV, 200 GeV) as compared to the other three combinations which we have
mentioned above. The stringent lower limits on $M$ in Figs. 1(b) and 1(c)
are about 550 GeV and 3190 GeV, respectively. From these observations
we can conclude that $BR(\mu\to e\gamma)$ has greater
sensitivity on $m_{\tilde l}$ as compared to that on $m_{\tilde \nu}$, and
lower the value of $m_{\tilde l}$ the greater would be the lower limit on
$M$. The lower bounds on $M$ increases with decreasing $v^\prime_1$,
since the elements of $Y_\nu$ will increase.
Another point to notice from the plots of Fig. 1 is that
$BR(\mu\to e\gamma)$ decreases with $M$ and goes to a dip at a certain value
of $M$, and then for a large value of $M$ it becomes saturate.
The reason for this is as follows. From the
amplitude of the process $\ell_j\to\ell_i+\gamma$, Eq. (\ref{E:ampli}),
we can notice that there is a relative minus sign between the contributions
of scalar and fermionic components of triplet Higgs. Moreover,
as explained before, in the numerical analysis, we have fixed the contribution
from scalar components by fixing their masses. Hence, due to the above mentioned relative
minus sign, at a certain value of $M$ the amplitude for
$\ell_j\to\ell_i+\gamma$ will become zero, and then goes to the saturation
for large value of $M$, since the amplitude is $\propto\frac{1}{M^2}$.
Since we have fixed the masses of scalar components of triplet Higgs to
the lower limits presented in Tab. 1, we here comment on what happens
if we increase their masses. Again, due to
the above mentioned relative minus sign, we can easily understand that the
lower bound on $M$ increases with $m_{\phi_1^{++}}$.
A final comment on the plots of Fig. 1 is that the bounds from the
decays $\tau\to e\gamma,\mu\gamma$ can be seen in the case of
$v^\prime_1$ = 0.1 eV but not in the cases of $v^\prime_1$ = 1.0 eV
and 0.5 eV. In Fig. 1(c) there are no points for $M<$ 450 GeV
and for $(m_{\tilde l},m_{\tilde \nu})$ = (200 GeV, 200 GeV), because
these points are are not satisfied by the experimental limits on
$Br(\tau\to e\gamma,\mu\gamma)$.

In Figs. 2(a) and 2(b) we have given constraints on $M$ in the
cases of IH and DN, respectively.
\begin{figure}[!h]
\begin{center}

\includegraphics[height=2.5in, width=2.5in]{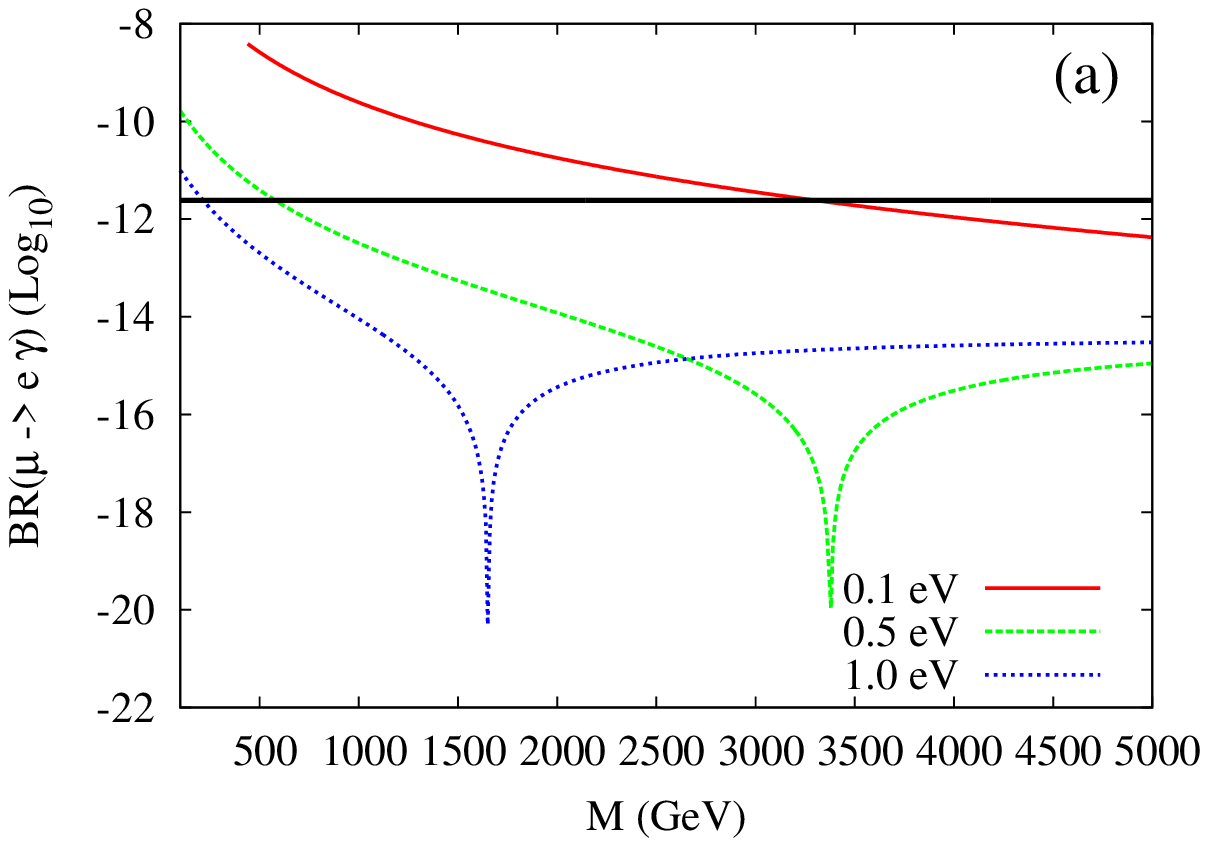}
\includegraphics[height=2.5in, width=2.5in]{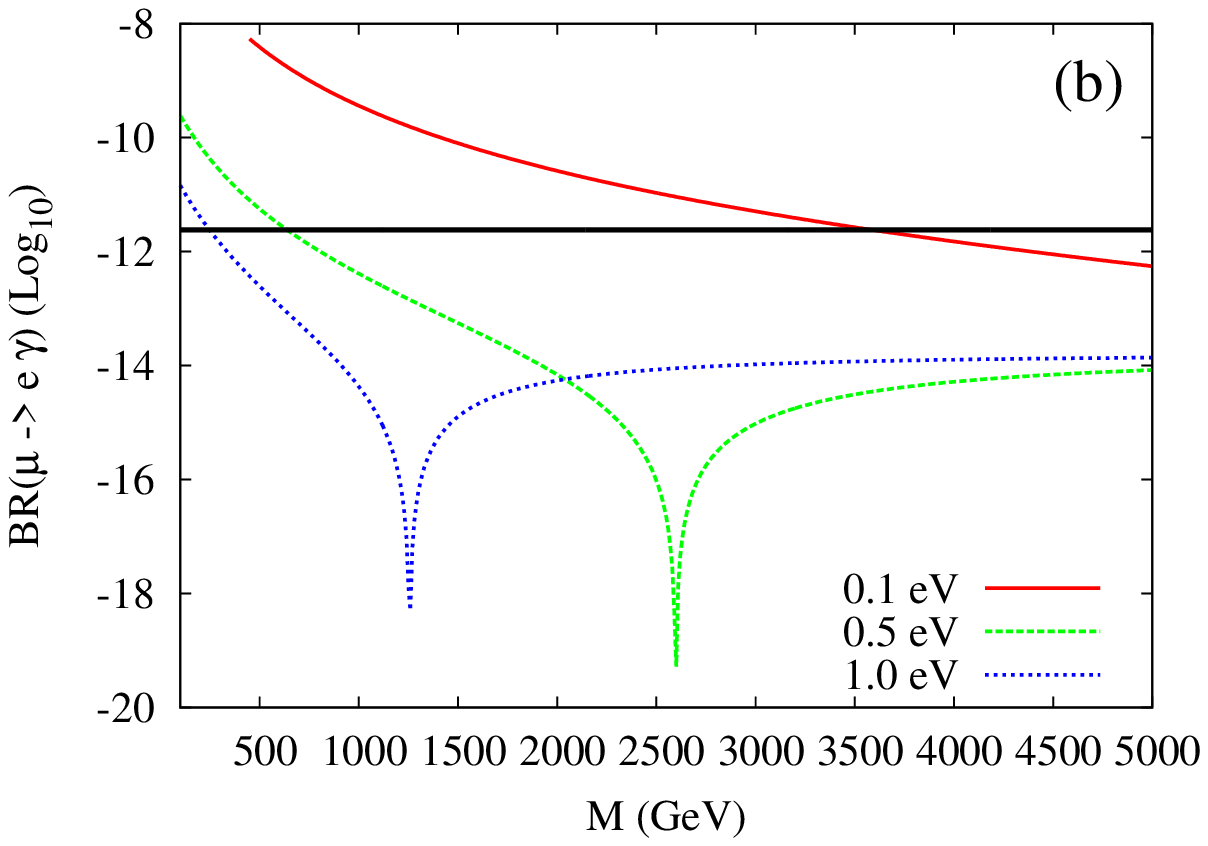}

\end{center}
\vspace*{-1cm}
\caption{In the left and right plots, $\log_{10}(BR(\mu\to e\gamma))$ has
been plotted against the mass of fermionic triplet Higgs,
in the cases of IH and DN, respectively.
In both of these plots, charged slepton and sneutrino masses have been taken
to be 200 GeV each. The three lines in each of the above plots are for different values of $v^\prime_1$
in eV units.
The horizontal line in these plots indicate $BR(\mu\to e\gamma)=2.4\times
10^{-12}$, and the area below this line is allowed.
The lower limit on the x-axis is 100 GeV.}
\end{figure}
In both of these cases, we have noticed that the dependence of
$BR(\mu\to e\gamma)$ on the $m_{\tilde l}$ and $m_{\tilde\nu}$ is same as that
described around Fig. 1. Hence, in both the plots of Fig. 2 we have fixed
$m_{\tilde l}$ and $m_{\tilde\nu}$ to a lower value of 200 GeV, which
should give stringent limits on $M$. We have varied $v^\prime_1$
in both the plots of Fig. 2. In the case of IH(DN) the lower
limits on $M$ for $v^\prime_1$ = 1.0 eV, 0.5 eV and 0.1 eV are
210(240) GeV, 570(630) GeV, 3300(3580) GeV, respectively. Comparing these
limits with the limits presented in the previous paragraph, the
lower bound on $M$ in the case of IH are intermediate between NH and
DN cases, and that the limits in the case of DN are stronger. In both of the
plots of Fig. 2, we can notice constraints arising from
$BR(\tau\to e\gamma, \mu\gamma)$ in the case of $v^\prime_1$ = 0.1 eV,
where points for $M<$ 450 GeV are not satisfied by them.

In Tab. 2, we have summarized the lower
limits on $M$ for different values of $v^\prime_1$ and in different
hierarchical mass patterns of neutrinos.
\begin{table}[!h]
\begin{center}
\begin{tabular}{||c|c|c|c||} \hline
 & NH & IH & DN \\ \hline
$v^\prime_1$ & $M$ & $M$ & $M$ \\
1.0 eV & 204 GeV & 216 GeV & 248 GeV \\
0.5 eV & 557 GeV & 579 GeV & 639 GeV \\
0.1 eV & 3.20 TeV & 3.31 TeV & 3.59 TeV \\ \hline
\end{tabular}
\end{center}
\caption{Lower bound on $M$ arising from
$BR(\mu\to e\gamma)<2.4\times 10^{-12}$,
for different values of $v^\prime_1$. These lower bounds are given in
all the three hierarchical mass patterns of neutrinos and for
$(m_{\tilde{l}},m_{\tilde{\nu}})$ = (200 GeV, 200 GeV). See text,
for more details.}
\end{table}
These lower bounds are given for $(m_{\tilde{l}},m_{\tilde{\nu}})$ =
(200 GeV, 200 GeV), in which case the limits on $M$ would be stringent.
Moreover,
while computing the lower bounds on $M$, we have fixed $m_{\phi_1^{++}}
\approx m_{\phi_1^+}$
to the values mentioned in Tab. 1.
By comparing the limits in Tab. 1 with that in Tab. 2, we can notice that the
lower bounds on $M$ are less than that on $m_{\phi^{++}_1}$. The reason
for this is that the bounds on $M$ and $m_{\phi^{++}_1}$ are coming
from LFV processes induced at 1-loop level and tree level, respectively.

After discussing the limits on the masses of scalar and fermionic
triplet Higgs states, which arise from LFV processes, we now discuss the
contribution of these triplet fields to the muon anomalous magnetic moment,
$(g-2)_\mu$ \cite{mug-2}. The current discrepancy between the SM
and the experimental value of $(g-2)_\mu$ can be taken as
$\Delta a_\mu = a_\mu^{\rm EXP} - a_\mu^{\rm SM} = (29\pm 9)\times 10^{-10}$ \cite{mug-2},
where $a_\mu=\frac{(g-2)_\mu}{2}$. The $(g-2)_\mu$ is a good observable quantity in
the study of new physics. In our model of SUSY Type II seesaw at TeV scale,
neutralino$-$charged slepton and chargino$-$sneutrino loops will give contribution
to the $(g-2)_\mu$ \cite{MSSMg-2}, and the above discrepancy can be easily
fitted.\footnote{For a recent fit to the $(g-2)_\mu$ in a model similar to the
MSSM, see Ref. \cite{Hundi}.} On top of this loop contribution, the
scalar and fermionic triplet Higgs states
will also give additional contribution to the $(g-2)_\mu$, which is
given in Eq. (\ref{E:g-2}). From the relation in Eq. (\ref{E:g-2}), we
can notice that the scalar and fermionic triplet Higgs states give
negative and positive contributions, respectively. Since the
current discrepancy in $\Delta a_\mu$ is strictly positive, the non-SUSY Type II
seesaw model, where the contribution is from scalar triplet Higgses,
cannot explain this discrepancy \cite{Fuku-etal}. In our present model, the fermionic
triplet Higgs states give positive contribution, so it is interesting to see
how large can this contribution be to the
$(g-2)_\mu$. The contribution of $\Delta a^{\rm T}_\mu$, Eq. (\ref{E:g-2}),
greatly depends on the sizes of Yukawa couplings. As mentioned before,
in the cases of NH and IH, for $v^\prime_1$ = 0.1 eV the Yukawa couplings 
are $\sim 10^{-2}$. Since these couplings are very small,
we do not expect appreciable amount to $\Delta a^{\rm T}_\mu$, in the
cases of NH and IH. Whereas, in the case of DN, the diagonal and
off-diagonal Yukawa couplings are around 1.5 and $\sim 10^{-3}$, respectively,
for $v^\prime_1$ = 0.1 eV. Hence, at least from the diagonal
Yukawa coupling $Y_\nu^{22}$
we can expect an enhancement to the $\Delta a^{\rm T}_\mu$.

In Fig. 3 we have plotted $\Delta a^{\rm T}_\mu$ versus
$M$ in the case of DN.
\begin{figure}[!h]
\begin{center}

\includegraphics[height=2.5in, width=2.5in]{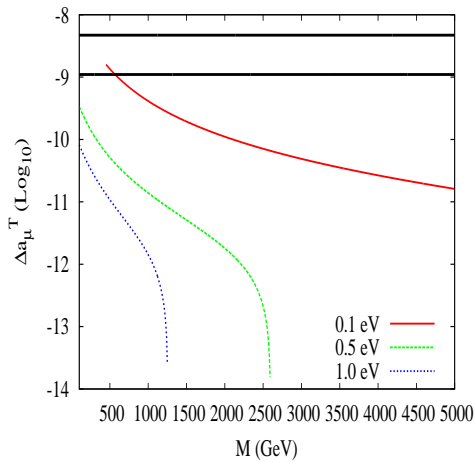}

\end{center}
\vspace*{-1cm}
\caption{The contribution of triplet Higgs states to
$\log_{10}(\Delta a_\mu^{\rm T})$
has been plotted against $M$, in the
case of DN. The masses of charged slepton and sneutrino have been
fixed to 200 GeV each. The three lines in this plot are for different values
of $v^\prime_1$, expressed in eV units. The horizontal lines represent the lower and upper
end of the $2\sigma$ limit of $\Delta a_\mu$, see text for details.}
\end{figure}
In this figure, we have kept the masses of scalar triplet
Higgses to the lower limits of Tab. 1, and also included the
constraints from $\tau\to e\gamma,\mu\gamma$.
The lower and upper horizontal lines in Fig. 3 represent the 2$\sigma$ limits
of the discrepancy in $\Delta a_\mu$, which can be taken as $1.1\times 10^{-9}$
and $4.7\times 10^{-9}$, respectively. The area between these lines is allowed
from the $(g-2)_\mu$. For $v^\prime_1$ = 1.0 eV and 0.5 eV the Yukawa couplings are
so small that the discrepancy in the $(g-2)_\mu$ cannot be fitted by the triplet
Higgses. Whereas for $v^\prime_1$ = 0.1 eV, there is a chance to fit this discrepancy
for a low value of $M$. However, the constraint from $\mu\to e\gamma$ puts a
lower limit on $M$ to be around 3600 GeV. Hence, after including the
constraints from LFV processes the maximum contribution to
the $(g-2)_\mu$ from triplet Higgses in this model is found to be $3.4\times 10^{-11}$
for $v^\prime_1$ = 0.1 eV, or 0.5 eV, or 1.0 eV. This contribution is two
orders smaller than the required amount. Hence, in the SUSY Type II
seesaw model, the discrepancy in $(g-2)_\mu$
can be fitted with the loop induced diagrams of
neutralino-charged slepton and chargino-sneutrino.
The reason for discontinuity of lines in Fig. 3 is that
after a certain large
value of $M$ the scalar contribution to $\Delta a^{\rm T}_\mu$ will be
dominant which is negative, and
we have plotted $\Delta a^{\rm T}_\mu$ in the units of $\log_{10}$.
The amount of this negative value
is so small that it gives negligible contribution to the $(g-2)_\mu$.

\section{Decays of scalar triplet Higgses}

The detection of components of triplet Higgs at the LHC can give
validity to our model. At the LHC or an $e^+e^-$ collider,
through the $\gamma$ and $Z$ mediated processes, both the charged as
well as the neutral components of triplet Higgses can be pair produced.
The production process for fermionic triplet Higgs states ($\Delta$s) at
a collider experiment is
similar to the corresponding production of charginos of the MSSM.
For the production of scalar triplet Higgses at collider
experiments, see Refs. \cite{scalpro,Aker-Aoki,Han-Muk-etal}.
Here, we study the decay products of
scalar triplet Higgses, through which the detection of these
fields can be done at collider experiments.
The decays of fermionic triplet Higgs states in a left-right
SUSY model can be found in Ref. \cite{Demir-etal}.

Among the scalar components of the triplet Higgs, decays of $\phi^{++}_1$,
$\phi^+_1$ and $\phi^0_1$ are interesting to study, since these states
are analogs of scalar triplet states in the non-SUSY version of Type II
seesaw model at TeV scale. As mentioned before that for $B_T\neq 0$, the
above mentioned $\phi_1$s will have mixing with the $\phi_2$s, and hence
$\phi_2$s can decay in the same way as $\phi_1$s do. Apart
from this, from gauge couplings and from $D$-terms in the SUSY scalar potential (see Appendix A), there
can also be decays like $\phi^{++}_{1,2}\to\phi^+_{2,1}W^+$,
$\phi^{++}_{1,2}\to\phi^{+}_{1,2} W^+$,
$\phi^{++}_{1,2}\to\phi^+_{1,2} H^+$, etc,
where $H^+$ is the charged component of the doublet
Higgs boson.
To simplify the many possible decays of scalar triplet Higgses, we
choose $B_T=0$ in our study here, which forbids decays of the form
$\phi^{++}_{1,2}\to\phi^+_{2,1}W^+$.
Also, as explained before, for $B_T=0$ mass splittings among various
charged components of $\phi_1$ and $\phi_2$ can be at most $\sim$ 10 GeV.
Hence, decays of the form $\phi^{++}_{1,2}\to\phi^+_{1,2} H^+$,
$\phi^{++}_{1,2}\to\phi^{+}_{1,2} W^+$, etc are kinematically
forbidden.
After this simplification is done, we can examine the distinction between
non-SUSY and SUSY versions of Type II seesaw model by studying the
decay patterns of $\phi_1$s.

The scalar $\phi_1$ states can decay into charged leptons and
neutrinos, and the interaction terms for these processes
can be read out from
Eq. (\ref{E:lagII}). The components of $\phi_1$
can also decay into scalar states containing charged sleptons and sneutrinos.
These decays are driven by the $(A_\nu Y_\nu)$-term of Eq. (\ref{E:vsoft}).
From the gauge invariant kinetic term of
the superfield $T_1$, the $\phi_1$s can
also decay into supersymmetric fields, whose
interaction terms can be obtained from
\begin{eqnarray}
{\cal L} &=& \left(T_1^\dagger e^{2gT^aW^a+2g^\prime B}T_1\right)_D,
\nonumber \\
&\ni&-\sqrt{2}\left(\phi^{++}_1\right)^*\left[(g\tilde{W}^3+g^\prime\tilde{B})\Delta^{++}_1
+g\tilde{W}^+\Delta^+_1\right]
-\sqrt{2}\left(\phi^{+}_1\right)^*\left[g\tilde{W}^-\Delta^{++}_1
+g^\prime\tilde{B}\Delta^+_1+g\tilde{W}^+\Delta^0_1\right]
\nonumber \\
&& -\sqrt{2}\left(\phi^{0}_1\right)^*\left[g\tilde{W}^-\Delta^+_1+
(-g\tilde{W}^3+g^\prime\tilde{B})\Delta^{0}_1\right]+{\rm h.c.}.
\end{eqnarray}
Here, $T^a$ are generators of the SU(2)$_L$
group in the triplet representation, which are given in Appendix A.
According to this representation, the form of $T_1$ in the above
equation should be $T_1=(T_1^{++},T_1^+,T_1^0)^{\rm T}$. $W^a,B$
are gauge superfields of the SU(2)$_L$ and U(1)$_Y$ groups,
respectively.
$\tilde{W}^\pm=\frac{1}{\sqrt{2}}(\tilde{W}^1\mp i\tilde{W}^2)$. In the
above equation, terms involving $\tilde{B}$ and $\tilde{W}^3$ give interactions
with the neutralinos, $N_k$, $k=1,\cdots,4$. Similarly, terms containing
$\tilde{W}^\pm$ give interactions with the charginos, $\chi^\pm_k$, $k=1,2$.
Our convention for the neutralino and chargino mass matrices and their
diagonalizing unitary matrices are given in Appendix B. In Appendix B,
we have taken $V^N$ and $V^\chi, U^\chi$ as the diagonalizing unitary
matrices for
neutralino and chargino mass matrices, respectively.

The decay widths of $\phi_1$s into leptonic and into SUSY fermionic
particles will have the following form.
\begin{eqnarray}
&&\Gamma(\phi_1\to AB)=\frac{1}{8\pi m_{\phi_1}^3}C_{\phi_1,A,B}
\sqrt{\lambda(m_{\phi_1},m_A,m_B)}(m_{\phi_1}^2-m_A^2-m_B^2),
\nonumber \\
&&\lambda(m_{\phi_1},m_A,m_B)=m_{\phi_1}^4+m_A^4+m_B^4-2m_{\phi_1}^2m_A^2
-2m_A^2m_B^2-2m_{\phi_1}^2m_B^2.
\label{E:dec1}
\end{eqnarray}
Whereas, the decay widths of $\phi_1$s into a pair of scalar states involving
charged sleptons or sneutrinos will have the following form.
\begin{equation}
\Gamma(\phi_1\to AB)=\frac{1}{16\pi m_{\phi_1}^3}C_{\phi_1,A,B}
\sqrt{\lambda(m_{\phi_1},m_A,m_B)}.
\label{E:dec2}
\end{equation}
Here, $A$ and $B$ are the product particles with masses $m_A$ and $m_B$,
respectively. $m_{\phi_1}$ is the mass of the parent particle $\phi_1$.
In the above Eqs. (\ref{E:dec1}) and (\ref{E:dec2}), the factor $C_{\phi_1,A,B}$ depends on the coupling
strength of the parent particle to the product particles, whose expressions are
given in Tab. 3.
\begin{table}[!h]
\begin{center}
\begin{tabular}{||cc||cc||} \hline
$\phi_1\to AB$ & $C_{\phi_1,A,B}$ & $\phi_1\to AB$ & $C_{\phi_1,A,B}$ \\ \hline
$\phi^{++}_1\to\ell^+_j\ell^+_k$ & $2S|Y_\nu^{jk}|^2$ &
$\phi^{++}_1\to\tilde{\ell}^+_j\tilde{\ell}^+_k$ & $S|(A_\nu Y_\nu)^{jk}|^2$ \\
$\phi^{++}_1\to\Delta^{++}_1N_k$ & $|gV^N_{2k}+g^\prime V^N_{1k}|^2$ &
$\phi^{++}_1\to\Delta^{+}_1\chi^+_k$ & $g^2|V^\chi_{1k}|^2$ \\ \hline
$\phi^{+}_1\to\nu_j\ell^+_k$ & $|Y_\nu^{jk}|^2$ &
$\phi^{+}_1\to\tilde{\nu}^*_j\tilde{\ell}^+_k$ & $\frac{1}{2}|(A_\nu Y_\nu)^{jk}+(A_\nu Y_\nu)^{kj}|^2$ \\
$\phi^{+}_1\to\Delta^{+}_1N_k$ & $g^{\prime 2}|V^N_{1k}|^2$ &
$\phi^{+}_1\to\Delta^{++}_1\chi^-_k$ & $g^2|U^\chi_{1k}|^2$ \\
$\phi^{+}_1\to\Delta^{0}_1\chi^+_k$ & $g^2|V^\chi_{1k}|^2$ & & \\ \hline
$\phi^{0}_1\to\nu_j\nu_k$ & $2S|Y_\nu^{jk}|^2$ &
$\phi^{0}_1\to\tilde{\nu}^*_j\tilde{\nu}^*_k$ & $S|(A_\nu Y_\nu)^{jk}|^2$ \\
$\phi^{0}_1\to\Delta^{0}_1N_k$ & $|gV^N_{2k}-g^\prime V^N_{1k}|^2$ &
$\phi^{0}_1\to\Delta^{+}_1\chi^-_k$ & $g^2|U^\chi_{1k}|^2$ \\ \hline
\end{tabular}
\end{center}
\caption{Various decay modes of $\phi_1$s and the factors $C_{\phi_1,A,B}$,
which are needed
in Eqs. (\ref{E:dec1}) and (\ref{E:dec2}). In the decay modes into leptons
and into sleptons,
$S$ is a symmetric factor which equals to $\frac{1}{2}$ if $j=k$, otherwise it equals
to 1.}
\end{table}

In this work we have considered the dominant tree level decays of
triplet scalar fields and have neglected loop induced decay processes.
At the tree level, there
can also be additional decays of $\phi_1$s into: (i) di-gauge bosons, (ii) a pair of
third family SM fermions, (iii) a pair involving components of doublet Higgses, (iv) gauge
boson and a component of doublet Higgs. Some of
the representative processes of these additional decays are as follows:
$\phi^{++}_1\to W^+W^+$, $\phi^+_1\to t\bar{b}$, $\phi^0_1\to H^+H^-$,
$\phi^+_1\to W^+H^0$. Except the decays in the category of (i), the decays in
(ii)$-$(iv) are driven due to the mixing between doublet and triplet scalar
Higgses \cite{Aker-Sugi}. However, coupling strengths of all the decays
in (i)$-$(iv) are proportional to $v^\prime_1$, which in our case is
very small, and hence the branching ratios of
these decays are negligible. Due to this,
we have neglected the above mentioned decays in our analysis.

The decay widths for $\phi_1$s into SUSY fermionic particles depend on their
SUSY masses as well as their coupling strengths, which can be uniquely determined
by the following set of parameters: $M_1$, $M_2$, $\mu$ and $\tan\beta$.
Here, $M_{1,2}$ are the soft masses of $\tilde{B}$ and $\tilde{W}^a$ fields,
respectively. In this work, we have chosen these
parameters as: $M_1=$ 200 GeV, $M_2=$ 300 GeV, $\mu =$ 400 GeV, $\tan\beta =$ 10.
This set of parameters give neutralino masses as: 195 GeV, 275 GeV, 405 GeV,
434 GeV, and the same set of parameters fix the chargino masses as: 274 GeV and 433 GeV.
The above choice of parameters is only for illustration. The qualitative
conclusions on
the branching ratios of $\phi_1$s do not change much with a different set of values.
We also have to fix the parameters $(A_\nu Y_\nu)^{jk}$ which drive the
decays of $\phi_1$s into charged sleptons and sneutrinos. For simplicity,
we take $(A_\nu Y_\nu)^{jk}=A_\nu(Y_\nu)^{jk}$ and we fix $A_\nu =$ 500 GeV.
As for the masses of charged sleptons and sneutrinos, we keep their masses to 200 GeV
each.

Below we have presented branching ratios of $\phi_1$s in the case of
NH. The choice of mass
spectrum of neutrinos fix the Yukawa couplings, which drive the decays
of $\phi_1$s into leptons and into sleptons, and this would effect the overall coefficients
of their branching ratios. Hence the qualitative features of the branching
ratios of $\phi_1$s would be similar in the other cases of IH and DN.

Decay modes of the scalar field $\phi^{++}_1$ are as follows: same sign charged dilepton
($\ell^+_j\ell^+_k$), same sign charged di-slepton ($\tilde{l}^+_j\tilde{l}^+_k$),
doubly charged fermionic triplet and neutralino
($\Delta^{++}_1N_k$), singly charged fermionic triplet and chargino ($\Delta^+_1\chi^+_k$).
The branching ratios of $\phi^{++}_1$ as function of its mass, in the case of
NH, are given in Fig. 4.
\begin{figure}[!h]
\begin{center}

\includegraphics[height=2.5in, width=2.5in]{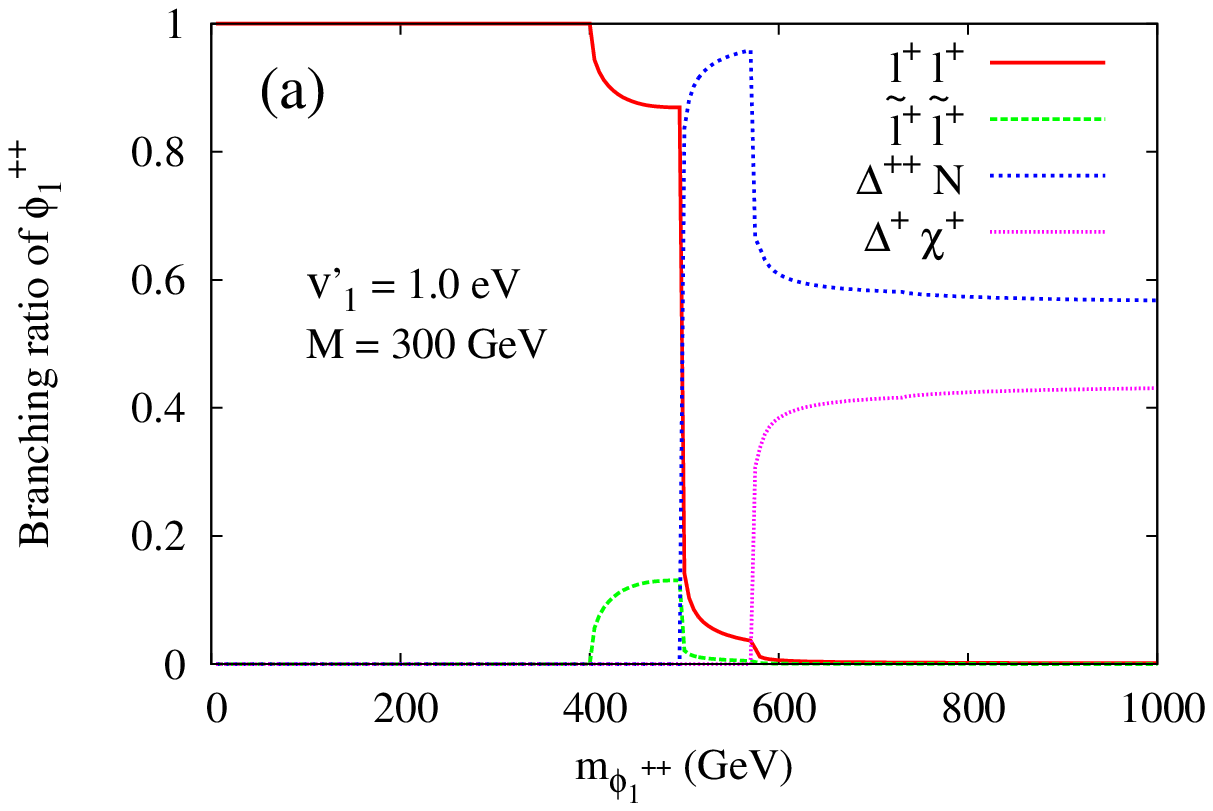}
\includegraphics[height=2.5in, width=2.5in]{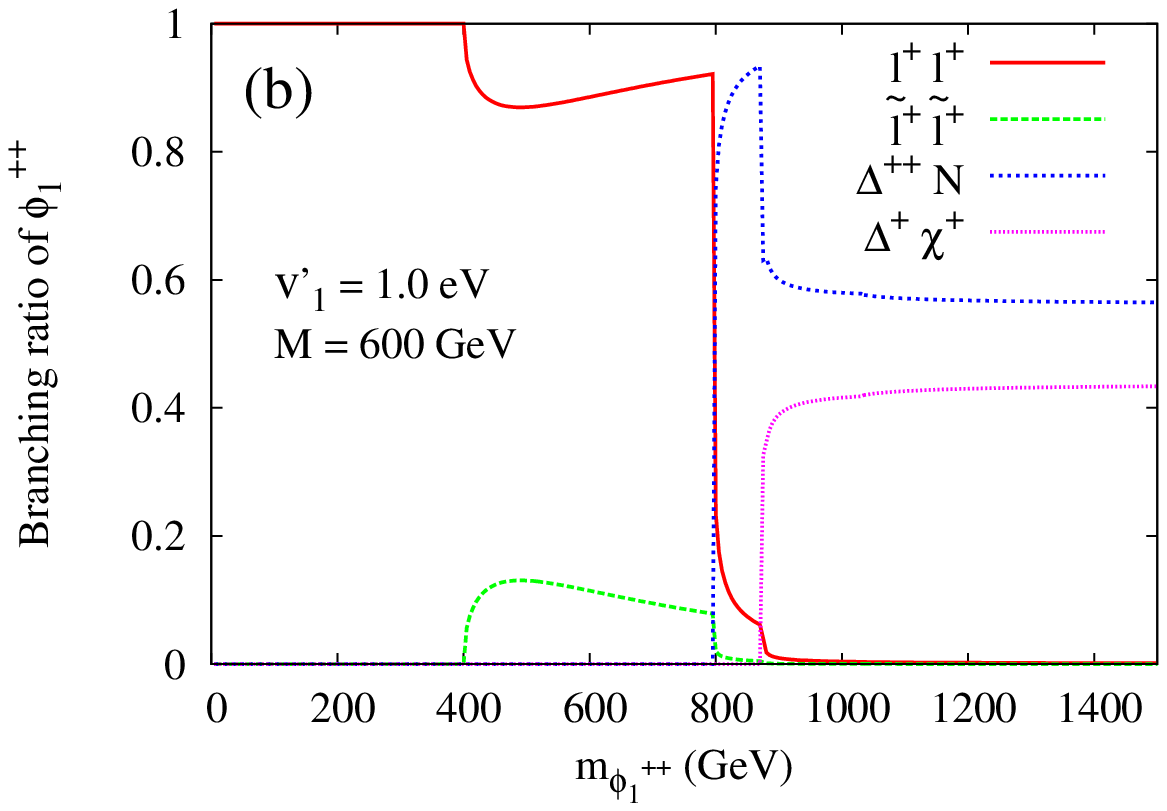}

\includegraphics[height=2.5in, width=2.5in]{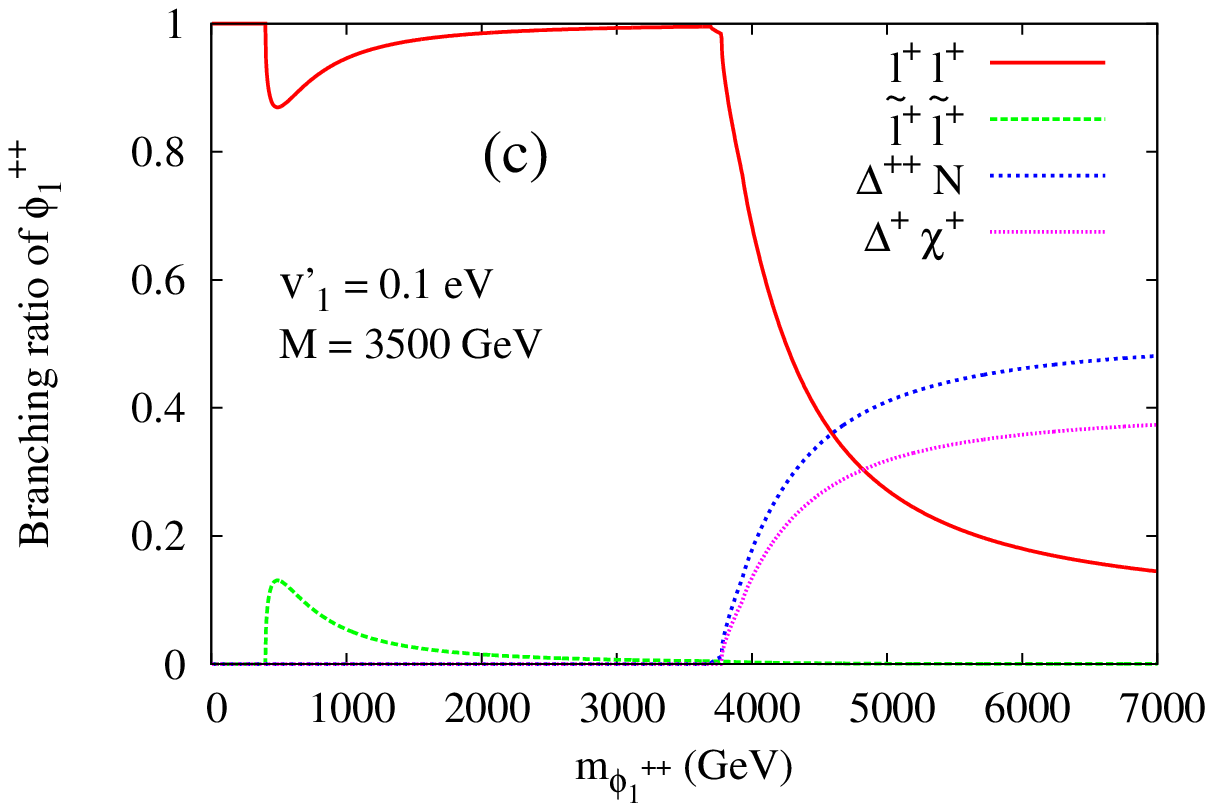}

\end{center}
\vspace*{-1cm}
\caption{Branching ratios of $\phi^{++}_1$ decay modes. In all
the decay modes we have summed over the generation index of product
particles, see text for details.}
\end{figure}
While plotting the branching ratios, we have summed over the indices
$j,k$. For instance, the branching ratio of $\phi^{++}_1$ into
same sign charged dileptons is taken as $Br(\phi^{++}_1\to\ell^+\ell^+)=
\frac{\sum_{j,k =1}^3\Gamma(\phi^{++}_1\to\ell^+_j\ell^+_k)}{\Gamma_{\phi^{++}_1}}$,
where $\Gamma_{\phi^{++}_1}$ is the total decay width of $\phi^{++}_1$.
Similarly, the three charged sleptons, the four neutralinos and the two charginos are summed
in the decay modes of $\phi^{++}_1\to\tilde{l}^+\tilde{l}^+$,
$\phi^{++}_1\to\Delta^{++}_1N$ and
$\phi^{++}_1\to\Delta^+_1\chi^+$, respectively. In the plots of Fig. 4, we
can notice that both the dilepton and
di-slepton modes will be suppressed as soon as the modes into SUSY fermionic
particles are kinematically accessible. The reason for this is as follows. Apart from
coupling strengths, in the limit of large mass of $\phi^{++}_1$,
the decay widths of $\phi^{++}_1$ into charged dilepton and into
SUSY fermionic particles vary as $\sim m_{\phi^{++}_1}$, while the corresponding
decay width for $\phi^{++}_1$ into charged di-slepton is $\sim\frac{A_\nu^2}{m_{\phi^{++}_1}}$.
From the above forms of decay widths, in the limit
$m_{\phi^{++}_1}\to \infty$, it is clear that
the decay mode into charged di-slepton cannot stand against decay modes
into dilepton and 
into SUSY fermionic particles. The decay modes into charged dileptons are driven
by Yukawa couplings, which are about $\sim 10^{-3}$ for $v^\prime_1=$ 1.0 eV.
Here the Yukawa couplings are far less than the gauge couplings
which drive the decay modes into SUSY fermionic particles, and hence these
modes are dominant over the charged dileptons.

In Fig. 4(a) we have chosen $v^\prime_1=$ 1.0 eV and the mass of fermionic
triplet is 300 GeV
which satisfies the flavor constraints described in the previous section.
For $M=$ 300 GeV, the SUSY modes involving the neutralinos and charginos
are kinematically accessible at about $m_{\phi^{++}_1}\sim$ 500 and 575 GeV, respectively.
As argued in the previous section, the LFV processes have
put a lower bound on $m_{\phi^{++}_1}$ to be about 630 GeV. Hence,
in the case of Fig. 4(a), the scalar field $\phi^{++}_1$ can be detected
in a collider experiment through its decays into SUSY fermionic particles,
because both the charged dilepton and charged di-slepton modes are suppressed for
$m_{\phi^{++}_1}>$ 630 GeV. However, in Fig. 4(b) we have increased
$M$ to 600 GeV so that the SUSY fermionic modes are kinematically accessible at about
$m_{\phi^{++}_1}\sim$ 800 GeV. Now, in this case, there is an appreciable branching
ratio of $\sim$ 90$\%$ to detect $\phi^{++}_1$ in the charged dilepton mode for
$m_{\phi^{++}_1}$ between about 630 to 800 GeV. In the same mass range of
$m_{\phi^{++}_1}\sim$ 630$-$800 GeV, the probability of
detecting $\phi^{++}_1$ in the charged di-slepton mode is hardly about 10$\%$.
However, by increasing $A_\nu$ from 500 GeV to 1 TeV, this probability can
be enhanced to 30$\%$, while at the same time the probability into
the charged dilepton mode will decrease to about 70$\%$. In Fig. 4(c) we have
decreased $v^\prime_1$ to 0.1 eV and have taken $M=$ 3500 GeV. In this
case, there will be enhancement in the Yukawa couplings compared to the previous
cases, and the lower limit on $m_{\phi^{++}_1}$ from the LFV processes
is about 6300 GeV. Because of the enhancement of the Yukawa couplings, the
decay mode into charged dilepton is still significant with a branching ratio of
$\sim 17\%$ for $m_{\phi^{++}_1}>$ 6300 GeV.

We can compare the results of Fig. 4
with that in the non-SUSY version of Type II seesaw model at TeV scale.
In the non-SUSY version, only the decay modes into dilepton and di-gauge boson
will be present \cite{Aker-Sugi}. However, as argued previously, the decay mode into
di-gauge boson will be suppressed in our context. The best
channel to detect a scalar triplet Higgs is in the decay $\phi^{++}_1
\to\ell^+\ell^+$, which has less background in a collider experiment.
However, in this model, this channel is restricted by the decay modes
into SUSY particles as well as by
constraints from the LFV processes. Whereas, in the non-SUSY version
of Type II seesaw model, even after imposing the constraints from
LFV processes, due to non-existence of decay modes into SUSY particles,
we would still have high branching ratio for the decay $\phi^{++}_1
\to\ell^+\ell^+$, provided $v^\prime_1<$ 0.1 MeV \cite{Aker-Sugi}.

Decay modes of the scalar field $\phi^+_1$ are as follows: neutrino
and charged lepton ($\nu_j\ell^+_k$), anti-sneutrino and charged slepton
($\tilde{\nu}^*_j\tilde{l}^+_k$), singly charged fermionic triplet
and neutralino ($\Delta^+_1N_k$), doubly charged fermionic triplet and
chargino ($\Delta^{++}_1\chi^-_k$), neutral fermionic triplet and
chargino ($\Delta^0_1\chi^+_k$). The branching ratios
of $\phi^+_1$ as function of its mass are given in Fig. 5, in the case of NH.
\begin{figure}[!h]
\begin{center}

\includegraphics[height=2.5in, width=2.5in]{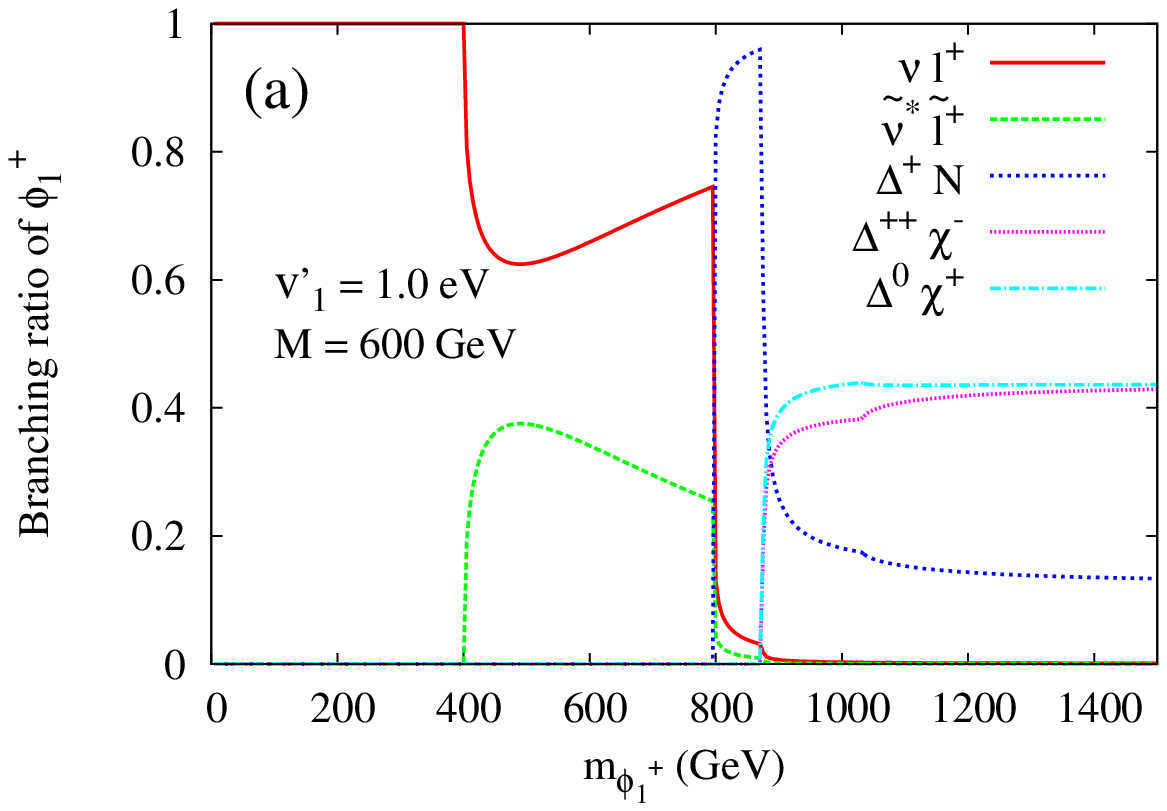}
\includegraphics[height=2.5in, width=2.5in]{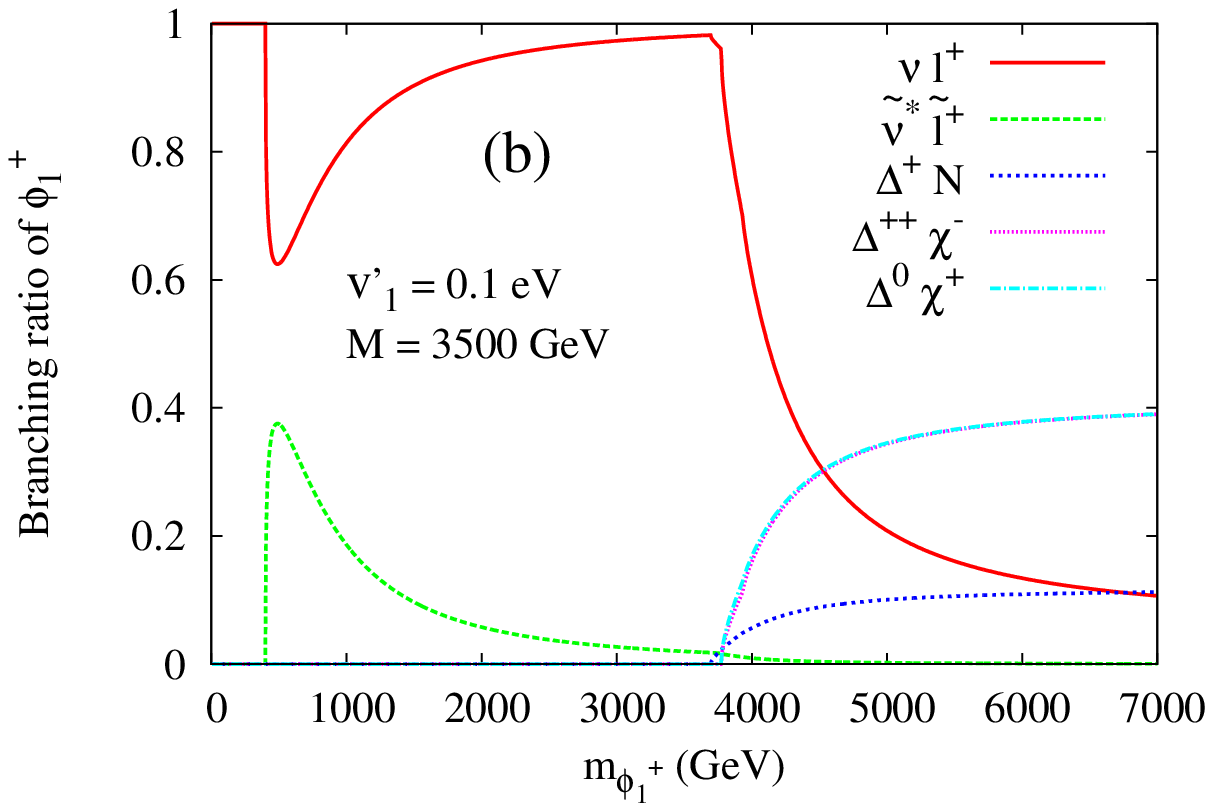}

\end{center}
\vspace*{-1cm}
\caption{Branching ratios of $\phi^{+}_1$ decay modes. In all
the decay modes we have summed over the generation index of product
particles, see text for details.}
\end{figure}
As explained around Fig. 4, here also, in the branching ratios of
$\phi_1^+$ into
leptons and into supersymmetric particles, we have summed over the indices $j,k$
of the leptons, sleptons, neutralinos and charginos. Like what happened in
the case of $\phi^{++}_1$ decays, in Fig. 5 we can observe that
the decay modes into leptons and into sleptons cannot stand
against the modes into SUSY fermionic particles. Unlike in the case of
$\phi^{++}_1$, the decay $\phi^+_1\to\nu\ell^+$ which is driven
by Yukawa couplings is not useful for detecting the scalar triplet, since
the neutrino is hard to detect in a collider experiment. Hence for
detecting the scalar field $\phi^+_1$, the modes into SUSY fermionic
particles are the best ones. The decay channel into $\tilde{\nu}^*\tilde{l}^+$
can be used for the
detection $\phi_1^+$ only for a certain choice of parametric values. In
Fig. 5(a) where $v^\prime_1=$ 1.0 eV and $M=$ 600 GeV, the
decay mode into $\tilde{\nu}^*\tilde{l}^+$
can be detected in the experiments with a branching ratio of nearly
30$\%$ for $m_{\phi^{+}_1}\sim$ 630$-$800 GeV. However, as explained
around Fig. 4, by decreasing $M$ below about 450 GeV and for
$v^\prime_1=$ 1.0 eV, the decay channel into
$\tilde{\nu}^*\tilde{l}^+$ would be suppressed.
In Fig. 5(b) we have taken $v^\prime_1=$
0.1 eV and $M=$ 3500 GeV. In this plot both the decay modes involving
chargino particles give approximately the same branching ratio.
By comparing the plots between Figs. 4 and 5,
we can notice that for a large value of $m_{\phi_1}$, the branching
ratio of $\phi_1\to\Delta N$ has higher value compared
to that of $\phi_1\to\Delta \chi$ in Fig. 4, whereas it is vice-versa in Fig. 5.
We believe the reason for this is that the coupling of $\phi^{++}_1(\phi^+_1)$
to $\Delta^{++}_1N_k(\Delta^+_1N_k)$ is proportional to
$gV^N_{2k}+g^\prime V^N_{1k}(g^{\prime}V^N_{1k})$.
Since $g>g^\prime$, that would explain the above mentioned observation.

Decay modes of the scalar field $\phi^0_1$ are as follows: pair of neutrinos
($\nu_j\nu_k$), pair of anti-sneutrinos ($\tilde{\nu}^*_j\tilde{\nu}^*_k$),
neutral fermionic triplet and neutralino
($\Delta^0_1N_k$), singly charged fermionic triplet and chargino
($\Delta^+_1\chi^-_k$).
The branching ratios of $\phi^0_1$ as function of its mass are
given in Fig. 6, in the case of NH.
\begin{figure}[!h]
\begin{center}

\includegraphics[height=2.5in, width=2.5in]{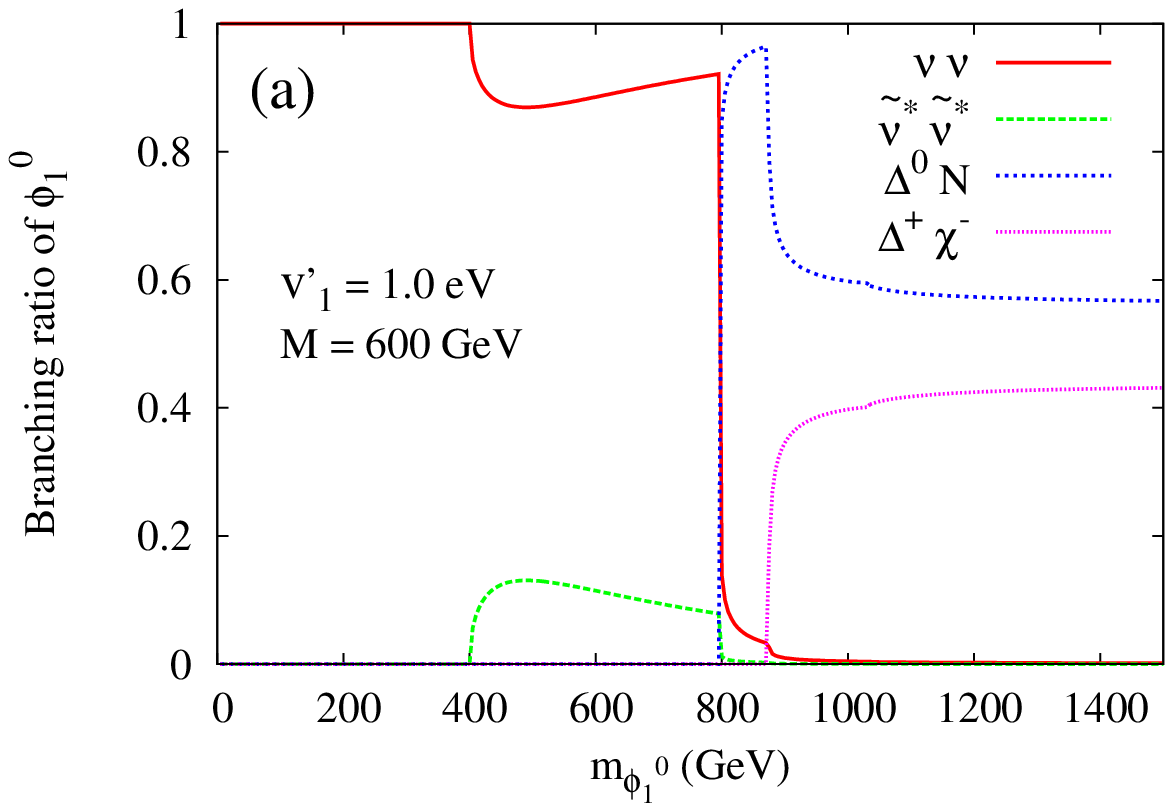}
\includegraphics[height=2.5in, width=2.5in]{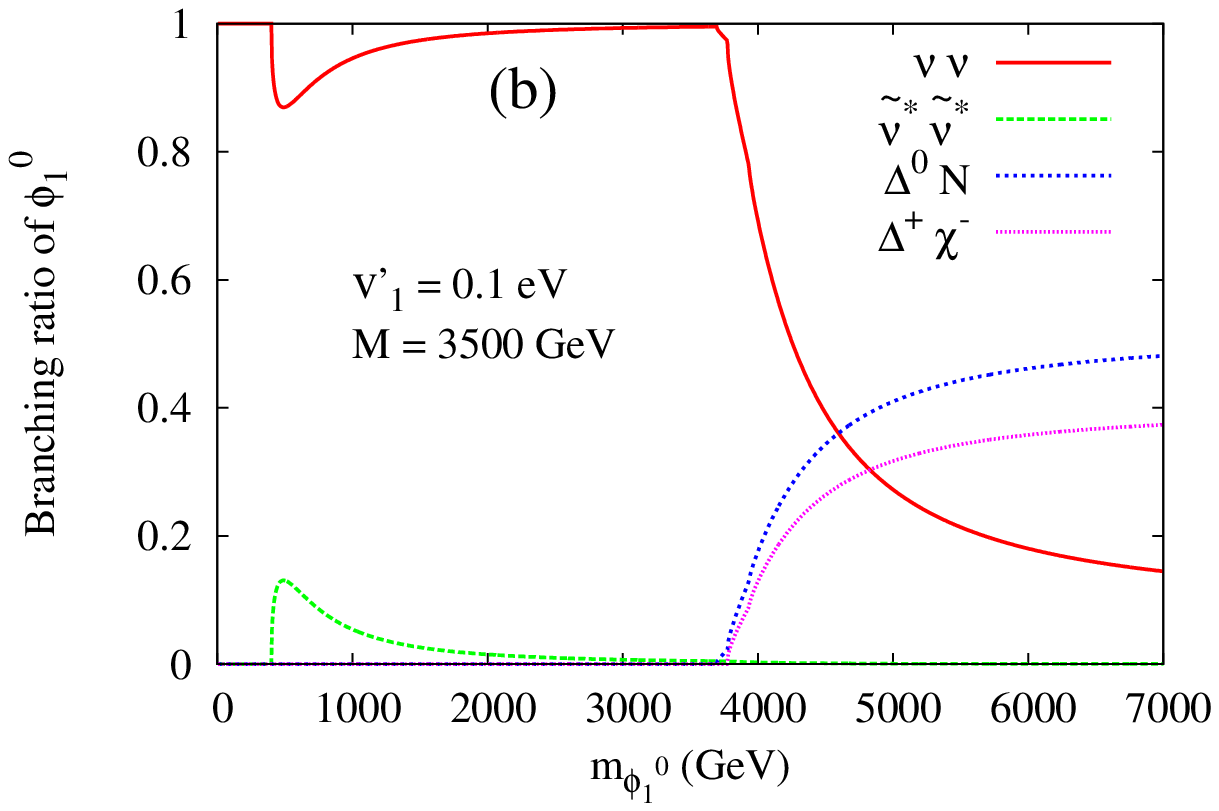}

\end{center}
\vspace*{-1cm}
\caption{Branching ratios of $\phi^{0}_1$ decay modes. In all
the decay modes we have summed over the generation index of product
particles, see text for details.}
\end{figure}
Similar to what we have done in Figs. 4 and 5, here also we have
summed over the indices $j,k$. As in the case for $\phi_1^+$,
the detection of $\phi^0_1$ can be mainly found from its
decays into SUSY fermionic particles. Also, for some specific choices of
$v^\prime_1$ and $M$, we can use the decay channel into a pair of
anti-sneutrinos for the detection of $\phi^0_1$. In Fig. 6(a),
the branching ratio for the decay channel into $\tilde{\nu}^*\tilde{\nu}^*$
is not larger than 10$\%$ in the allowed region of $m_{\phi^0_1}\sim$
630$-$800 GeV. However, this branching ratio can be increased by increasing
the value of $A_\nu$ from its input value of 500 GeV.

A final comment on the decay branching ratios of $\phi^+_1$ and $\phi^0_1$,
which are described in Figs. 5 and 6, are as follows. In the littlest
Higgs model with SU(5) symmetry \cite{Arkani-etal}, both the doublet and
triplet scalar states of the gauged SU(2)$_L$ will be put into one single
SU(5) multiplet. As a result of this, we can see that
the decay branching ratios of $\phi^+\to t\bar{b}$ and
$\phi^0\to t\bar{t}$ are significant \cite{Han-etal}
in the littlest Higgs
model. However, as explained
before, in our model, the above mentioned
decay processes are suppressed due to small admixture between doublet
and triplet scalar fields. Hence, in a collider experiment, the decays of
$\phi_1^+$ and $\phi_1^0$ can be
used to distinguish the Type II seesaw model at TeV scale and the
littlest Higgs model.

So far we have dealt with the decays of $\phi_1$s and in this model there is
another scalar triplet $\phi_2$ with a hypercharge of $-1$. As explained before,
to simplify the decay channels of scalar triplets, we have chosen $B_T=0$.
As a result of this, the charged and neutral components of $\phi_2$ will
dominantly decay into the modes involving SUSY fermionic particles such as
fermionic triplet Higgses, neutralinos and charginos. The expressions for
the decay widths of these modes are
similar to the corresponding $\phi_1$ decay modes, Eq. (\ref{E:dec1}).
The decay modes of $\phi_2$ have no competing channels
involving dilepton, and hence the branching
ratios of $\phi_2$ will be constant in the limit of large masses of these
fields.

\section{Detection prospects of low energy SUSY Type II seesaw model}

As explained in the previous section, one way of probing the low energy
Type II seesaw model is to look for signals of scalar triplet Higgs fields
in collider experiments. The decay modes of scalar triplet fields which
are driven by the Yukawa couplings should be looked at collider experiments
in order to verify the neutrino mass mechanism. In this regard, the
decay channel $\phi_1^{++}\to\ell^+\ell^+$ is the best mode to probe in
experiments. At the LHC, doubly charged scalar triplets can be pair produced
through Drell-Yan process. It has been reported in Ref. \cite{Han-Muk-etal}
that the cross section for this pair production is about 1 to 0.1 fb for
$m_{\phi^{++}}$ between about 600 to 1000 GeV. In the case of Drell-Yan
process, the final signal would be 4 leptons of the form
$\ell^+\ell^+\ell^-\ell^-$. One can also singly produce $\phi^{\pm\pm}$ at the
LHC through the process $q^\prime\bar{q}\to W^*\to \phi^{\pm\pm}\phi^\mp$
\cite{Aker-Aoki}. It has been claimed in Ref. \cite{Aker-Aoki} that
the cross section for the single production of $\phi^{\pm\pm}$ can be
enhanced by about a factor of 2 compared to the Drell-Yan case. In the case of
single production of $\phi^{\pm\pm}$, the final signal would be 3 leptons of the
form $\ell^\pm\ell^\pm\ell^\mp$.
In the previous section, we have described that the decay modes of
scalar triplets into leptons will compete 
with decay modes into supersymmetric particles.
In Figs. 4(b) and 5(a), $\phi^{++}_1\to\ell^+\ell^+$ and $\phi^+_1\to\ell^+\nu$
have appreciable branching ratios of $\sim$0.9 and $\sim$0.7, respectively,
for $m_{\phi^{++}_1}$ between 630 to 800 GeV.
Assuming an
integrated luminosity of 100 fb$^{-1}$ at the LHC in future, we can
observe about 8 to 80 events for $m_{\phi^{++}_1}$ between 630 to 800 GeV,
in the case of 4-lepton
signal. In the case of 3-lepton signal, about 12 to 120 events can be
observed at the LHC. However, these event numbers are calculated without
including background processes and simulation cuts, and a detailed
analysis should be done in order to detect the scalar triplet fields
at the LHC. Apart from the above described 4-lepton and 3-lepton signals,
there can be other possibilities in our model. In either of the
processes $q\bar{q}(\bar{q}^\prime)\to\phi^{++}\phi^{--}(\phi^-)$,
one doubly charged scalar triplet can decay into dilepton, whereas,
the other scalar triplet can decay into SUSY particles. In these
processes, there can be
flavor violating decays $\phi^{--}\to e\mu,\mu\tau$,
etc at the LHC.

We comment on the detection prospects of our model compared to the
Type II seesaw models where the triplet fields are super heavy. As already
described before, in our model the triplet fields have masses around
1 TeV and we have assumed that off-diagonal entries in the soft masses
of sleptons are zero. However, in models where triplet fields are
super heavy, due to renormalization effects, slepton mass matrix can acquire
non-zero off-diagonal elements.
In fact, in this class of models
\cite{Rossi,heavyTpII}, it has been shown
that various LFV processes are correlated by
the same model parameters, and flavor violating decays of staus and
neutralinos can be observed at the LHC \cite{heavyTpII}.
In our model these processes are
absent, however, LFV decays of charged triplet fields
of this model can be observed at the LHC. We have commented
on one such possibility in the previous paragraph.

In our model the flavor
violation is driven by the off-diagonal elements of Yukawa couplings,
$Y_\nu$. Hence, flavor violation in our model can be probed at LHC
in the decay modes of $\phi_1^{++}$ into charged leptons, which
are presented in Fig. 4. Since the Yukawa couplings
in different hierarchical patterns of neutrinos would be different,
$BR(\phi_1^{++}\to\ell\ell)$ would be different in these different cases,
which should offer different signal strength
at the LHC.
Moreover, as described before, the elements of $Y_\nu$ are nearly
$\sim 10^{-3}$ for $v^\prime_1$ = 1 eV in the cases of NH and IH. Whereas,
in the case of DN, the diagonal and off-diagonal elements of $Y_\nu$
are $\sim$0.1 and $\sim 10^{-4}$, respectively, for $v^\prime_1$ = 1 eV.
The difference in Yukawa couplings give different values for, say
$BR(\phi_1^{++}\to e^+e^+)$ and $BR(\phi_1^{++}\to e^+\mu^+)$, which
should be probed at LHC to distinguish the case of DN from NH and IH.

Apart from accelerator based experiments, neutrinoless double beta decay
($0\nu\beta\beta$) experiments offer alternative probes for new physics models
where neutrinos
are Majorana particles. In the Type II seesaw model at TeV scale, the scalar
triplet Higgses, through a sub-process $W^-W^-\to\phi^{--}\to\ell^-\ell^-$,
can give additional contribution to $0\nu\beta\beta$ \cite{Rode}.
But due to small couplings and heavy masses of these fields,
this additional contribution is highly negligible, and hence
the amplitude for $0\nu\beta\beta$ is dominantly contributed by the
Majorana neutrinos \cite{Rode}.

\section{Conclusions}

In this work we have focused on the phenomenological implications of
supersymmetric Type II seesaw model at TeV scale. In this model, there are
two triplet superfields with hypercharges $Y=+1,-1$, whose scalar
and fermionic components will have masses at around TeV scale. Also,
the smallness of neutrino masses can be naturally explained in this
model, provided
the vevs of the neutral scalar triplet fields are around 1 eV. In this
scenario, the Yukawa couplings of the triplet field ($Y=+1$) to the
lepton doublets are unsuppressed and these couplings can drive LFV
processes. We have focused on a particular parameter space of the model
where the loop induced processes due to charged slepton and sneutrino
fields give negligible contribution
to the LFV processes. Another simplified assumption
we have made is that we have neglected the mixing between scalar components
of the two different triplet superfields.

After making the above assumptions, the branching ratios of LFV processes
in the Type II seesaw model depend dominantly on the Yukawa couplings
and masses of
the triplet fields. The Yukawa couplings in this model are
determined by the neutrino oscillation data and the scalar triplet
vev $v_1^\prime$.
Specifically, we have found that among the various possible
LFV processes, the current experimental upper limits on the branching ratios
of $\mu\to 3e$ and $\mu\to e\gamma$ can put lower limits on the
masses of triplet Higgs states.
The masses of scalar ($m_\phi$) and fermionic ($M$) triplet Higgs states should
be at least 630 and 200 GeV, respectively. We have
tabulated the lower limits on the masses of these fields in Tabs. 1 and 2.
The lower limits on $m_\phi$ and $M$ depend on the hierarchical
mass pattern of neutrinos and also on $v_1^\prime$. The bounds
on $M$ also depend on the masses of sleptons.
For slepton masses as low as 200 GeV we get stringent bounds on $M$,
which are displayed in Tab. 2.

Next, we have addressed the implications of the constraints from LFV
processes on observable quantities such as the muon anomalous magnetic moment,
$(g-2)_\mu$. In the case of degenerate mass pattern of neutrinos,
the contribution to the $(g-2)_\mu$ from the scalar and fermionic triplet
Higgses can fit the current discrepancy in it. However, after
applying the constraints from the above described LFV processes,
this contribution will be at most $3.4\times 10^{-11}$, which is two
orders less than the required amount.

We have also studied the detection of scalar triplet fields
in a collider experiment, and for this
we have studied decay patterns of these fields. While studying these decay
processes, we have applied the same assumptions which we have applied
in our study on the LFV processes, which are described above. As a
result of this, the scalar triplet fields ($\phi_1$s) which have
hypercharge $Y=+1$
can decay into leptonic as well as supersymmetric particles. Whereas,
the scalar triplet fields which have hypercharge $Y=-1$ can decay only
into supersymmetric fields.
The golden channel to detect any of these
scalar triplet Higgses is $\phi^{++}_1\to\ell^+\ell^+$, and we
have addressed how this channel will be affected due to the presence
of decay modes involving supersymmetric particles. Our study suggests
that the above mentioned golden channel may not compete against the
decay modes into supersymmetric particles. However, for some suitable
choice of model parameters, where $M$ = 600 GeV and for $m_\phi$
between about 630 to 800 GeV, $BR(\phi^{++}_1\to\ell^+\ell^+)$ can be
as high as 90$\%$. Similarly, after applying the constraints from LFV
and depending on the choice of parameter space, the singly charged
($\phi^+_1$) and neutral ($\phi^0_1$) scalar triplet Higgses can
be detected in the modes involving supersymmetric fields.

Flavor violation in our model can take place through decays such as
$\phi^{--}\to e\mu,\mu\tau,$ etc. Probing such flavor violation in
the LHC can not only test the Type II seesaw mechanism
of our model but
also can be used to distinguish different hierarchical mass patterns of
neutrinos.

\section*{Acknowledgments}

The author is thankful to Dilip Kumar Ghosh for valuable discussions
and also for reading the manuscript.

\section*{Appendix}

\subsection*{A) Scalar potential}

The scalar potential of the SUSY Type II seesaw model at TeV scale will
have the following form.
\begin{eqnarray}
V&=&\sum_Y\left|\frac{\partial W}{\partial Y}\right|^2+\frac{1}{2}\sum_{a=1}^3D^aD^a
+\frac{1}{2}D^YD^Y
+V_{\rm soft}^{\rm MSSM}+V_{\rm soft}^{\rm triplet},
\label{E:V1} \\
D^a&=&-g\left(H_d^\dagger\frac{\sigma^a}{2}H_d+H_u^\dagger\frac{\sigma^a}{2}H_u+\Phi_1^\dagger T^a\Phi_1
+\Phi_2^\dagger T^a\Phi_2\right),
\nonumber \\
D^Y&=&-\frac{g^\prime}{2}\left(H^\dagger_uH_u-H^\dagger_dH_d\right)-g^\prime
\left(\Phi_1^\dagger\Phi_1-\Phi^\dagger_2\Phi_2\right).
\nonumber
\end{eqnarray}
The first term in Eq. (\ref{E:V1}) is the $F$-term contribution where the
summation over the fields $Y$ run over the superfields of $W$ of Eq. (\ref{E:WII}).
The second and third terms of Eq. (\ref{E:V1}) are $D$-term contributions
due to SU(2)$_L$ and U(1)$_Y$ gauge groups, respectively. The last two terms
of Eq. (\ref{E:V1}) are soft terms of MSSM and of fields involving triplet
scalar fields. The form of $V_{\rm soft}^{\rm MSSM}$ can be found in
Ref. \cite{susy,Martin}. The $V_{\rm soft}^{\rm triplet}$ is given in Eq. (\ref{E:vsoft}).
The triplet representation of SU(2) generators, which are needed in $D^a$
of Eq. (\ref{E:V1}), are
\begin{equation}
T^1=\frac{1}{\sqrt{2}}
\left(\begin{array}{ccc}
0 & 1 & 0 \\
1 & 0 & 1 \\
0 & 1 & 0
\end{array}\right), \quad
T^2=\frac{1}{\sqrt{2}}
\left(\begin{array}{ccc}
0 & -i & 0 \\
i & 0 & -i \\
0 & i & 0
\end{array}\right), \quad
T^3=
\left(\begin{array}{ccc}
1 & 0 & 0 \\
0 & 0 & 0 \\
0 & 0 & -1
\end{array}\right).
\end{equation}
For computing $D$-terms, the forms of the scalar fields are as follows: $H_u=(H^+_u,H^0_u)^{\rm T}$,
$H_d=(H^0_d,H^-_d)^{\rm T}$, $\Phi_1=(\phi_1^{++},\phi^+_1,\phi^0_1)^{\rm T}$,
$\Phi_2=(\phi_2^{0},\phi^-_2,\phi^{--}_2)^{\rm T}$. We have described the
$D$-terms of doublet and of triplet Higgses, but the $D$-terms for other scalar fields
of the model can be analogously written.

\subsection*{B) Conventions of neutralino and chargino mass matrices and their
diagonalizing matrices}

Our conventions regarding neutralino and chargino mass matrices are same
as in \cite{Martin}.
In the basis $\Psi^0=(\tilde{B},\tilde{W}^3,\tilde{H}_d,\tilde{H}_u)^{\rm T}$,
the mixing mass matrix of neutralinos can be written as ${\cal L}_N=
-\frac{1}{2}\left(\Psi^0\right)^{\rm T}M_N\Psi^0+{\rm h.c.}$. The form of
$M_N$ is same as Eq. (8.2.2) of Ref. \cite{Martin}.
The physical neutralino states are defined from $\Psi^0_j=\sum_{k=1}^4V^N_{jk}N_k$, where
the unitary matrix $V^N$ diagonalizes $M_N$ as
\begin{equation}
\left(V^N\right)^{\rm T}M_NV^N={\rm diag}(m_{N1},m_{N2},m_{N3},m_{N4}).
\end{equation}

In the basis: $\Psi^-=(\tilde{W}^-,\tilde{H}^-_d)^{\rm T}$,
$\Psi^+=(\tilde{W}^+,\tilde{H}^+_u)^{\rm T}$, the mixing mass terms
for charginos can be written as ${\cal L}_c=-\left(\Psi^-\right)^{\rm T}
M_C\Psi^++{\rm h.c.}$. The matrix $M_C$ is same as Eq. (8.2.14)
of Ref. \cite{Martin}.
The physical chargino states are defined from: $\Psi^-_j =\sum_{k=1}^2
U^\chi_{jk}~\chi^-_k$, $\Psi^+_j=\sum_{k=1}^2V^\chi_{jk}~\chi^+_k$. The
unitary matrices $U^\chi$ and $V^\chi$ diagonalize $M_C$ as
\begin{equation}
\left(U^\chi\right)^{\rm T}M_CV^\chi ={\rm diag}(m_{C1},m_{C2}).
\end{equation}


\begin{thebibliography}{99}

\bibitem{LHC-Higgs}
  G.~Aad {\it et al.}  [ATLAS Collaboration],
    Phys.\ Lett.\ B {\bf 716}, 1 (2012)
    [arXiv:1207.7214 [hep-ex]];
  S.~Chatrchyan {\it et al.}  [CMS Collaboration],
    Phys.\ Lett.\ B {\bf 716}, 30 (2012)
    [arXiv:1207.7235 [hep-ex]].

\bibitem{BSM}
  M.~E.~Peskin,
  In *Carry-le-Rouet 1996, High-energy physics* 49-142
  [hep-ph/9705479];
  G.~Altarelli,
  Nucl.\ Instrum.\ Meth.\ A {\bf 518}, 1 (2004)
  [hep-ph/0306055];
  C.~Quigg,
  hep-ph/0404228;
  J.~Ellis,
  Nucl.\ Phys.\ A {\bf 827}, 187C (2009)
  [arXiv:0902.0357 [hep-ph]].

\bibitem{fcnc-rev}
  T.~Mori,
    eConf C {\bf 060409}, 034 (2006)
    [hep-ex/0605116];
  J.~M.~Yang,
    Int.\ J.\ Mod.\ Phys.\ A {\bf 23}, 3343 (2008)
    [arXiv:0801.0210 [hep-ph]];
  A.~J.~Buras,
    Acta Phys.\ Polon.\ Supp.\  {\bf 3}, 7 (2010)
    [arXiv:0910.1481 [hep-ph]];
  Y.~Nir,
  CERN Yellow Report CERN-2010-001, 279-314
  [arXiv:1010.2666 [hep-ph]].

\bibitem{neut-rev}
For a review on neutrino masses and mixing, see
  R.~N.~Mohapatra,
  hep-ph/0211252;
  Y.~Grossman,
  hep-ph/0305245;
  A.~Strumia and F.~Vissani,
    hep-ph/0606054.

\bibitem{Tp-II}
  M.~Magg and C.~Wetterich,
  Phys.\ Lett.\  B {\bf 94}, 61 (1980);
  J.~Schechter and J.~W.~F.~Valle,
  Phys.\ Rev.\ D {\bf 22}, 2227 (1980);
  R.~N.~Mohapatra and G.~Senjanovic,
  Phys.\ Rev.\  {\bf D23}, 165 (1981);
  G.~Lazarides, Q.~Shafi and C.~Wetterich,
  Nucl.\ Phys.\  B {\bf 181}, 287 (1981).

\bibitem{susy}
  H.~P.~Nilles, Phys. Rept. {\bf 110}, 1 (1984);
  H.~E.~Haber and G.~L.~Kane, Phys. Rept. {\bf 117}, 75 (1985);
  M.~Drees, R.~Godbole and P.~Roy, Theory and Phenomenology of Sparticles,
  (World Scientific, 2004);
  P.~Binetruy, Supersymmetry (Oxford University Press, 2006);
  H.~Baer and X.~Tata, Weak Scale Supersymmetry: From
  Superfields to Scattering Events, (Cambridge University Press, 2006).

\bibitem{Martin}
  S.~P.~Martin, arXiv:hep-ph/9709356.

\bibitem{susyTp-II}
  T.~Hambye, E.~Ma and U.~Sarkar,
  Nucl.\ Phys.\ B {\bf 602}, 23 (2001)
  [hep-ph/0011192].

\bibitem{Rossi}
  A.~Rossi,
  Phys.\ Rev.\ D {\bf 66}, 075003 (2002)
  [hep-ph/0207006].

\bibitem{heavyTpII}
  M.~Hirsch, S.~Kaneko and W.~Porod,
  Phys.\ Rev.\ D {\bf 78}, 093004 (2008)
  [arXiv:0806.3361 [hep-ph]];
  J.~N.~Esteves, J.~C.~Romao, A.~Villanova del Moral, M.~Hirsch, J.~W.~F.~Valle and W.~Porod,
  JHEP {\bf 0905}, 003 (2009)
  [arXiv:0903.1408 [hep-ph]].

\bibitem{lepgen}
  S.~Antusch and S.~F.~King,
  Phys.\ Lett.\ B {\bf 597}, 199 (2004)
  [hep-ph/0405093];
  E.~J.~Chun and S.~Scopel,
  Phys.\ Lett.\ B {\bf 636}, 278 (2006)
  [hep-ph/0510170];
  M.~Senami and K.~Yamamoto,
  Int.\ J.\ Mod.\ Phys.\ A {\bf 21}, 1291 (2006)
  [hep-ph/0305202].

\bibitem{hgamgam}
  A.~G.~Akeroyd and S.~Moretti,
  Phys.\ Rev.\ D {\bf 86}, 035015 (2012)
  [arXiv:1206.0535 [hep-ph]].

\bibitem{pdg}
  J. Beringer {\it et al.} (Particle Data Group), Phys.\ Rev.\ D {\bf 86}, 010001 (2012).

\bibitem{TpIIwk}
  E.~J.~Chun, K.~Y.~Lee and S.~C.~Park,
  Phys.\ Lett.\ B {\bf 566}, 142 (2003)
  [hep-ph/0304069];
  M.~Kakizaki, Y.~Ogura and F.~Shima,
  Phys.\ Lett.\ B {\bf 566}, 210 (2003)
  [hep-ph/0304254];
  E.~K.~.Akhmedov and W.~Rodejohann,
  JHEP {\bf 0806}, 106 (2008)
  [arXiv:0803.2417 [hep-ph]];
  W.~Rodejohann,
  Pramana {\bf 72}, 217 (2009)
  [arXiv:0804.3925 [hep-ph]].

\bibitem{Aker-etal}
  A.~G.~Akeroyd, M.~Aoki and H.~Sugiyama,
  Phys.\ Rev.\ D {\bf 79}, 113010 (2009)
  [arXiv:0904.3640 [hep-ph]].

\bibitem{Fuku-etal}
  T.~Fukuyama, H.~Sugiyama and K.~Tsumura,
  JHEP {\bf 1003}, 044 (2010)
  [arXiv:0909.4943 [hep-ph]].

\bibitem{Sena-Yama}
  M.~Senami and K.~Yamamoto,
  Phys.\ Rev.\ D {\bf 69}, 035004 (2004)
  [hep-ph/0305203].

\bibitem{mug-2}
For a review on the muon $(g-2)$, see,
  Z.~Zhang,
  arXiv:0801.4905 [hep-ph];
  F.~Jegerlehner and A.~Nyffeler,
  Phys.\ Rept.\  {\bf 477}, 1 (2009)
  [arXiv:0902.3360 [hep-ph]].

\bibitem{HPT}
  R.~S.~Hundi, S.~Pakvasa and X.~Tata,
  Phys.\ Rev.\ D {\bf 79}, 095011 (2009)
  [arXiv:0903.1631 [hep-ph]].

\bibitem{MEG}
  J.~Adam {\it et al.}  [MEG Collaboration],
  Phys.\ Rev.\ Lett.\  {\bf 107}, 171801 (2011)
  [arXiv:1107.5547 [hep-ex]].

\bibitem{Chak-etal}
  J.~Chakrabortty, P.~Ghosh and W.~Rodejohann,
  arXiv:1204.1000 [hep-ph].

\bibitem{theta13}
  Y.~Abe {\it et al.}  [DOUBLE-CHOOZ Collaboration],
  Phys.\ Rev.\ Lett.\  {\bf 108}, 131801 (2012)
  [arXiv:1112.6353 [hep-ex]];
  F.~P.~An {\it et al.}  [DAYA-BAY Collaboration],
  Phys.\ Rev.\ Lett.\  {\bf 108}, 171803 (2012)
  [arXiv:1203.1669 [hep-ex]];
  J.~K.~Ahn {\it et al.}  [RENO Collaboration],
  Phys.\ Rev.\ Lett.\  {\bf 108}, 191802 (2012)
  [arXiv:1204.0626 [hep-ex]].

\bibitem{glob-fit}
  D.~V.~Forero, M.~Tortola and J.~W.~F.~Valle,
  arXiv:1205.4018 [hep-ph].

\bibitem{tribi}
  P.~F.~Harrison, D.~H.~Perkins and W.~G.~Scott,
  Phys.\ Lett.\ B {\bf 530}, 167 (2002)
  [hep-ph/0202074].

\bibitem{CMS}
  S.~Chatrchyan {\it et al.}  [CMS Collaboration],
  arXiv:1207.2666 [hep-ex].

\bibitem{MSSMg-2}
  T.~Moroi,
  Phys.\ Rev.\ D {\bf 53}, 6565 (1996)
  [Erratum-ibid.\ D {\bf 56}, 4424 (1997)]
  [hep-ph/9512396];
  S.~P.~Martin and J.~D.~Wells,
  Phys.\ Rev.\ D {\bf 64}, 035003 (2001)
  [hep-ph/0103067].

\bibitem{Hundi}
  R.~S.~Hundi,
  Phys.\ Rev.\ D {\bf 83}, 115019 (2011)
  [arXiv:1101.2810 [hep-ph]].

\bibitem{scalpro}
  J.~F.~Gunion, R.~Vega and J.~Wudka,
  Phys.\ Rev.\ D {\bf 42}, 1673 (1990);
  R.~Godbole, B.~Mukhopadhyaya and M.~Nowakowski,
  Phys.\ Lett.\ B {\bf 352}, 388 (1995)
  [hep-ph/9411324];
  K.~-m.~Cheung, R.~J.~N.~Phillips and A.~Pilaftsis,
  Phys.\ Rev.\ D {\bf 51}, 4731 (1995)
  [hep-ph/9411333];
  M.~Muhlleitner and M.~Spira,
  Phys.\ Rev.\ D {\bf 68}, 117701 (2003)
  [hep-ph/0305288];
  E.~J.~Chun and P.~Sharma,
  JHEP {\bf 1208}, 162 (2012)
  [arXiv:1206.6278 [hep-ph]].

\bibitem{Aker-Aoki}
  A.~G.~Akeroyd and M.~Aoki,
  Phys.\ Rev.\ D {\bf 72}, 035011 (2005)
  [hep-ph/0506176].

\bibitem{Han-Muk-etal}
  T.~Han, B.~Mukhopadhyaya, Z.~Si and K.~Wang,
  Phys.\ Rev.\ D {\bf 76}, 075013 (2007)
  [arXiv:0706.0441 [hep-ph]].

\bibitem{Demir-etal}
  D.~A.~Demir, M.~Frank, D.~K.~Ghosh, K.~Huitu, S.~K.~Rai and I.~Turan,
  Phys.\ Rev.\ D {\bf 79}, 095006 (2009)
  [arXiv:0903.3955 [hep-ph]].

\bibitem{Aker-Sugi}
  A.~G.~Akeroyd and H.~Sugiyama,
  Phys.\ Rev.\ D {\bf 84}, 035010 (2011)
  [arXiv:1105.2209 [hep-ph]].

\bibitem{Arkani-etal}
  N.~Arkani-Hamed, A.~G.~Cohen, E.~Katz and A.~E.~Nelson,
  JHEP {\bf 0207}, 034 (2002)
  [hep-ph/0206021].

\bibitem{Han-etal}
  T.~Han, H.~E.~Logan, B.~Mukhopadhyaya and R.~Srikanth,
  Phys.\ Rev.\ D {\bf 72}, 053007 (2005)
  [hep-ph/0505260].

\bibitem{Rode}
  W.~Rodejohann,
  Int.\ J.\ Mod.\ Phys.\ E {\bf 20}, 1833 (2011)
  [arXiv:1106.1334 [hep-ph]].

\end{thebibliography}
\end{document}